\documentclass[preprint2]{aastex}

\usepackage{amsmath}
\usepackage{graphicx}
\usepackage{subfigure}
\usepackage{natbib}

\newcommand{\rsun}{$R_\odot$}

\slugcomment{Submitted to ApJ on April 10, 2015. Accepted for publication on August 15, 2015.}

\shorttitle{Physical Conditions of Coronal Plasma at the transit of a Shock driven by a CME}
\shortauthors{Susino, Bemporad \& Mancuso}

\begin{document}


\title{Physical Conditions of Coronal Plasma at the transit of a Shock\\
    driven by a Coronal Mass Ejection}


\author{R. Susino\altaffilmark{1}}
\email{susino@oato.inaf.it}

\author{A. Bemporad\altaffilmark{1} and S. Mancuso\altaffilmark{1}}

\affil{INAF-Turin Astrophysical Observatory, via Osservatorio 20, 10025 Pino Torinese (TO), Italy}



\begin{abstract}
We report here on the determination of plasma physical parameters across a shock driven by a Coronal Mass Ejection using White Light (WL) coronagraphic images and Radio Dynamic Spectra (RDS). The event analyzed here is the spectacular eruption that occurred on June 7th 2011, a fast CME followed by the ejection of columns of chromospheric plasma, part of them falling back to the solar surface, associated with a M2.5 flare and a type-II radio burst. Images acquired by the SOHO/LASCO coronagraphs (C2 and C3) were employed to track the CME-driven shock in the corona between 2--12~\rsun\ in an angular interval of about 110$^\circ$. In these intervals we derived 2-Dimensional (2D) maps of electron density, shock velocity and shock compression ratio, and we measured the shock inclination angle with respect to the radial direction. Under plausible assumptions, these quantities were used to infer 2D maps of shock Mach number $M_\text{A}$ and strength of coronal magnetic fields at the shock's heights. We found that in the early phases (2--4~\rsun) the whole shock surface is super-Alfv\'enic, while later on (i.e. higher up) it becomes super-Alfvenic only at the nose. This is in agreement with the location for the source of the observed type-II burst, as inferred from RDS combined with the shock kinematic and coronal densities derived from WL. For the first time, a coronal shock is used to derive a 2D map of the coronal magnetic field strength over a 10~\rsun\ altitude and $\sim110^\circ$ latitude intervals.
\end{abstract}


\section{Introduction} \label{sec:intro}

The study of Interplanetary Shocks associated with major solar eruptions is very important  not only from the theoretical point of view, but also because of potential impacts on human technologies. First because shocks are, as well as solar flares, optimal locations for the acceleration of Solar Energetic Particles (SEPs; i.e. electrons, protons and He ions with energies from a few KeV to some GeV) that constitute an important hazard for satellites and astronauts, and may affect the ionosphere around polar caps. Moreover, as the shocks reach the Earth, significant southward components of the interplanetary magnetic field associated with them can magnetically reconnect with the magnetosphere, thus disturbing the system and producing severe geomagnetic storms \citep[see e.g. review by][]{schwenn2006}. Hence, understanding the origin, propagation and physical properties of interplanetary shocks is also crucial for future developments of our capabilities of forecasting possible Space Weather effects of solar activity. For these reasons, over the last decades huge efforts have been devoted in order to improve our knowledge of these phenomena and of the associated Coronal Mass Ejections (CMEs), by using different instrumentation taking remote sensing as well as in situ data. In particular, over the last few years, the most recent space based missions, such as the twin STEREO satellites, the Hinode and SDO observatories, provided significant new insights, thus allowing to investigate shocks from the early phases of their formation at the base of the corona out to their propagation into the interplanetary space.

A clear signature of the formation and propagation of interplanetary shocks associated with CME expansion and/or flare explosions is the detection of type-II radio bursts \citep[see][for a review of the problem of type-II sources]{vrsnak2008}. Combination of radio data with images acquired at different wavelengths is able to provide unique new information on these phenomena. Recently, combined analysis of EUV images and radio dynamic spectra were used to demonstrate \citep{cho2013,chen2014} that type-II bursts may be excited in the lower corona through interaction between CMEs and nearby dense structures such as streamers \citep[see also][]{classenaurass2002, reiner2003, ma04}. A similar result was also obtained with the use of a new radio triangulation technique exploiting radio data acquired by different spacecraft \citep{magdalenic2014}. Hence, type-II radio bursts are likely to be excited during the early propagation phase of the shocks (that is, at heliocentric distances $r < 1.5$~\rsun), around the expected location of the local minimum of $v_\text{A}(r)$ profile \citep{gopalswamy2012a, gopalswamy2013}. Thanks to the high cadence, good sensitivity and spatial resolution now available in EUV with SDO/AIA, it has been shown \citep{kouloumvakos2014} also that the sole analysis of EUV images can provide by itself an estimate of the density compression ratio $X$ (an important shock parameter given by the ratio between the downstream and the upstream plasma densities, $X = n_\text{d} / n_\text{u}$) and that this estimate is in agreement with the one derived from radio data in sheat regions. The above results clearly have important implications for the identification of SEP source regions.

Over the last decade it also became clear that a significant number of information on interplanetary shocks can be derived from White Light (WL) coronagraphs data alone, as first shown by \citet{vourlidas03}. Analysis of these data allowed to verify that shocks form when their propagation velocity $v_\text{sh}$ (measured in a reference system at rest with the solar wind, moving at velocity $v_\text{sw}$) is larger than the local Alfv\'en velocity $v_\text{A}$ ($|\mathbf{v_\text{sh}} - \mathbf{v_\text{sw}}| > v_\text{A} = B/ \sqrt{4 \pi \rho}$). Hence, the lower is the velocity of the driver, the larger are the distances where shock front forms \citep{eselevich2011}. Moreover, combination of EUV and WL data shows that the shock thickness $\delta$ is of the same order as the proton mean free path $\lambda_p$ only for heliocentric distances $r < 6$~\rsun\, while higher up in the corona $\delta << \lambda_p$. Hence, during its propagation, the shock regime changes from collisional to collisionless \citep{eselevich2012}. These information are crucial for our understanding of the physics at the base of the shock. Also, at larger heliocentric distances, the analysis of WL data provided by heliospheric imagers have demonstrated that the driver (CME) and the shock undergo different magnetic drag deceleration during their interplanetary expansion, with the shock propagating faster than the ejecta, thus leading to possible CME-shock decouplings \citep[][]{hesszhang2014}. Statistically, the coupling has been found to be stronger for faster CMEs \citep{mujiber2013}. Studies of interplanetary propagation of shocks have tremendous implications for Space Weather prediction capabilities as well.

\begin{figure*}
\centering
\includegraphics[width=\textwidth]{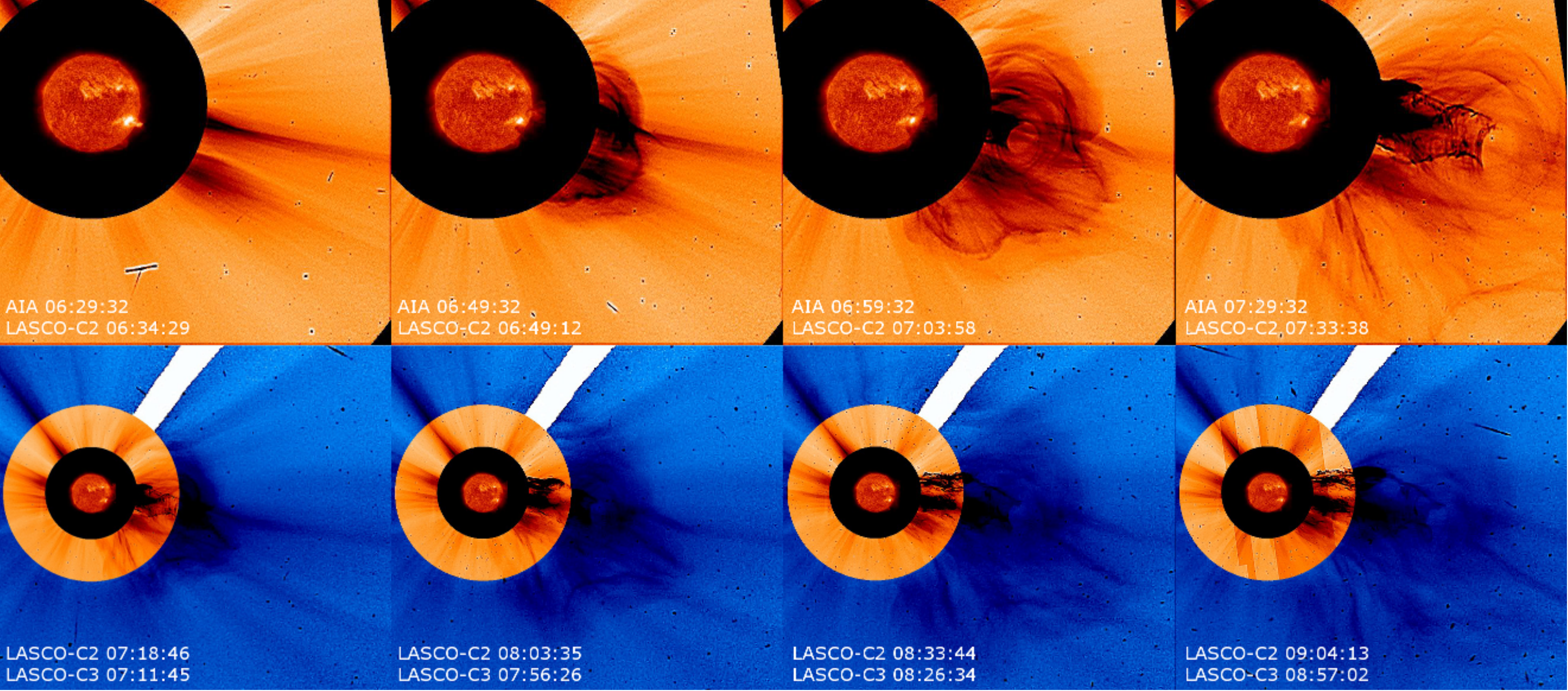}
\caption{Top: sequence of SDO/AIA 304 and SOHO/LASCO-C2 images acquired on June 7, 2011 during the eruptive event analyzed here. The LASCO-C2 images are shown in inverted color scale (brighter features are darker and vice-versa) and after the application of a filter to enhance the visibility of CME structures (images created with JHelioviewer). Bottom: sequence of LASCO-C2 and -C3 images showing the CME propagation at higher altitudes; again the images are shown in inverted color scale and after the application of a filter to enhance the visibility of CME structures (images created with JHelioviewer).}\label{fig:event}
\end{figure*}

Significant advances were also made from comparisons between observations and numerical simulations. At heliocentric distances $r > 2$~\rsun\, coronal protons and electrons are no more coupled by Coulomb collisions. This leads to different temperatures for these two species, with slightly larger proton than electron temperatures (by a factor depending on the relevant altitude and coronal structure) as demonstrated by coronal UV spectra acquired by the UV coronagraph Spectrometer \citep[UVCS; see reviews by][]{antonucci2006,kohl06}. Protons, however, being much heavier than electrons, have much smaller microscopic velocities (by a factor of 42.85). CME-driven shocks are thus supersonic only with respect to the proton thermal speed, implying that only protons are expected to be significantly heated by the transit of the shock. This was recently confirmed from both observations and simulations: in particular, \citet{manchester2012, jin2013} demonstrate that the WL appearances of CME-driven shocks are better reproduced by 2-temperature (2T) MHD simulations with respect to 1-temperature (1T) simulations, where 2T plasma protons are heated up to $\sim 90$~MK, and 2T shocks have larger Alfv\'enic Mach numbers $M_\text{A}$ (by a factor $\sim 1.25$--1.4) with respect to the 1T plasma case. Very similar results were recently obtained by the combined analysis of UV and WL observations of a CME driven shock performed by \citet{bem14}.

The latter work was the result of a sequence of previous researches performed on CME-driven shocks and based on the combined analyses of UV spectra acquired by UVCS and WL images acquired by the LASCO coronagraph. As first demonstrated by \citet{bem10}, this unique combination allows to measure not only the plasma compression ratio $X$, but also the pre- and post-shock plasma temperatures. Moreover, once these informations are combined with the Rankine-Hugoniot equations written for the general case of oblique shocks, and by measuring geometrical (inclination) and kinematical (velocity) properties of the shock from WL data, it is even possible to determine both the pre- and post-shock magnetic and velocity field vectors projected on the plane of the sky. This technique allowed \citet{bem11,bem13} to conclude that, for a few specific events, radio-loud (radio-quiet) CMEs are more likely associated with super- (sub-) critical shocks, and that only a small region around the shock center is super-critical in the early evolution phases, while higher up (i.e. later on) the whole shock becomes sub-critical. Moreover, the same technique applied to different points located along the same shock front allowed \citet{bem14} to demonstrate that the transit of shock leads to a significant deflection of the magnetic field close to the shock nose, and a smaller deflection at the flanks, implying a draping of field lines around the expanding CME, in nice agreement with the post-shock magnetic field rotations obtained by \citet{liu2011} with 3D MHD numerical simulations.

In this paper the above results are further extended: in particular we demonstrate here that, under some specific hypotheses, the analysis of WL coronagraphic data alone not only can provide the density compression ratios at different times and locations along the shock front, but also the $M_\text{A}$ numbers and the pre-shock coronal magnetic fields, allowing us to derive a 2D map of magnetic field strength covering an heliocentric distance interval by $\sim 10$~\rsun\ and a latitude interval by $\sim 110^\circ$. Moreover, the combined analysis of WL and radio data allows us to derive the possible location of the source for the type-II radio burst. The paper is organized as follows: after a general description of the event being analyzed here (Section \ref{sec:obs}), we describe the analysis of data (Section \ref{sec:datanal}), focusing in particular on LASCO/C2 and C3 WL coronagraphic images (Section \ref{sec:wldata}) and WAVES/RAD1-RAD2 radio dynamic spectra (Section \ref{sec:radiodata}). Then, the obtained results are summarized and discussed (Section \ref{sec:concl}).

\section{Observations} \label{sec:obs}

On June 7th 2011, a GOES M2.6 class flare from AR 11226 (located in the southwest quadrant at 22$^\circ$ S and 66$^\circ$ W) occurred between 06:16 and 06:59 UT, peaking around 06:16 UT. This soft X-ray flare was associated with significant HXR emission and even $\gamma-$ray emission lasting for about 2 hours \citep{ackermann2014}. The impressive eruption associated with this flare has been extensively studied by many previous authors who focused on different physical phenomena related with the event. They focused on several aspects of this event, such as the early evolution of the released CME bubble and compression front \citep{cheng2012}, the propagating EUV wave \citep{li2012}, the magnetic reconnections driven by the CME expansion \citep{vandriel2014}, the flare emission \citep{inglisgilbert2013}, and the associated type-II radio burst \citep{dorovskyy2013, dorovskyy2015}. Moreover, this spectacular eruption was followed by the ejection of huge radial columns of chromospheric plasma, reaching the field of view of LASCO and COR1 coronagraphs, and then falling back to the sun. Thus, other authors focused also on the dynamics and plasma properties of returning plasma blobs \citep{innes2012, williams2013, carlyle2014, dolei2014}, as well as on the energy release from falling material impact on the sun \citep{gilbert2013, reale2013, reale2014}.

In this work we study the evolution of the shock wave associated with this eruption as observed by white light coronagraphic images. As reported by \citet{cheng2012}, immediately after the flare onset (around 06:26 UT) a circular plasma CME bubble was observed in the SDO/AIA images expanding at $\sim 960$~km~s$^{-1}$; in the early phases, due to the small standoff distance, the compression front and the front of the driver (i.e. the CME bubble) cannot be discerned. The two fronts started to separate only later on, when a deceleration of the CME bubble is observed; at the same time, a type II radio burst started (as well as a type-III burst), suggesting that the compression wave had just turned itself into a shock wave. Later on, the CME enters in the field of view of the SOHO/LASCO-C2 coronagraph starting from the frame acquired at 06:49 UT (Figure 1, top row), and then enters in the field of view of the LASCO C3 coronagraph starting form the frame acquired at 07:11 UT (Figure 1, bottom row). The LASCO C2 frames clearly show the propagation of the shock wave associated with the event, as well as the CME front and the circular flux rope, while this latter part becomes hardly discernible in the LASCO C3 frames (see \ref{fig:event}).

In what follows we describe how the sequence of white light images acquired by LASCO C2 and C3 has been analyzed to derive the pre-CME coronal density and the different physical parameters of the shock wave.

\section{Data analysis} \label{sec:datanal}

\subsection{WL coronagraphic images} \label{sec:wldata}

\begin{figure*}
\centering
\includegraphics[width=0.7\textwidth]{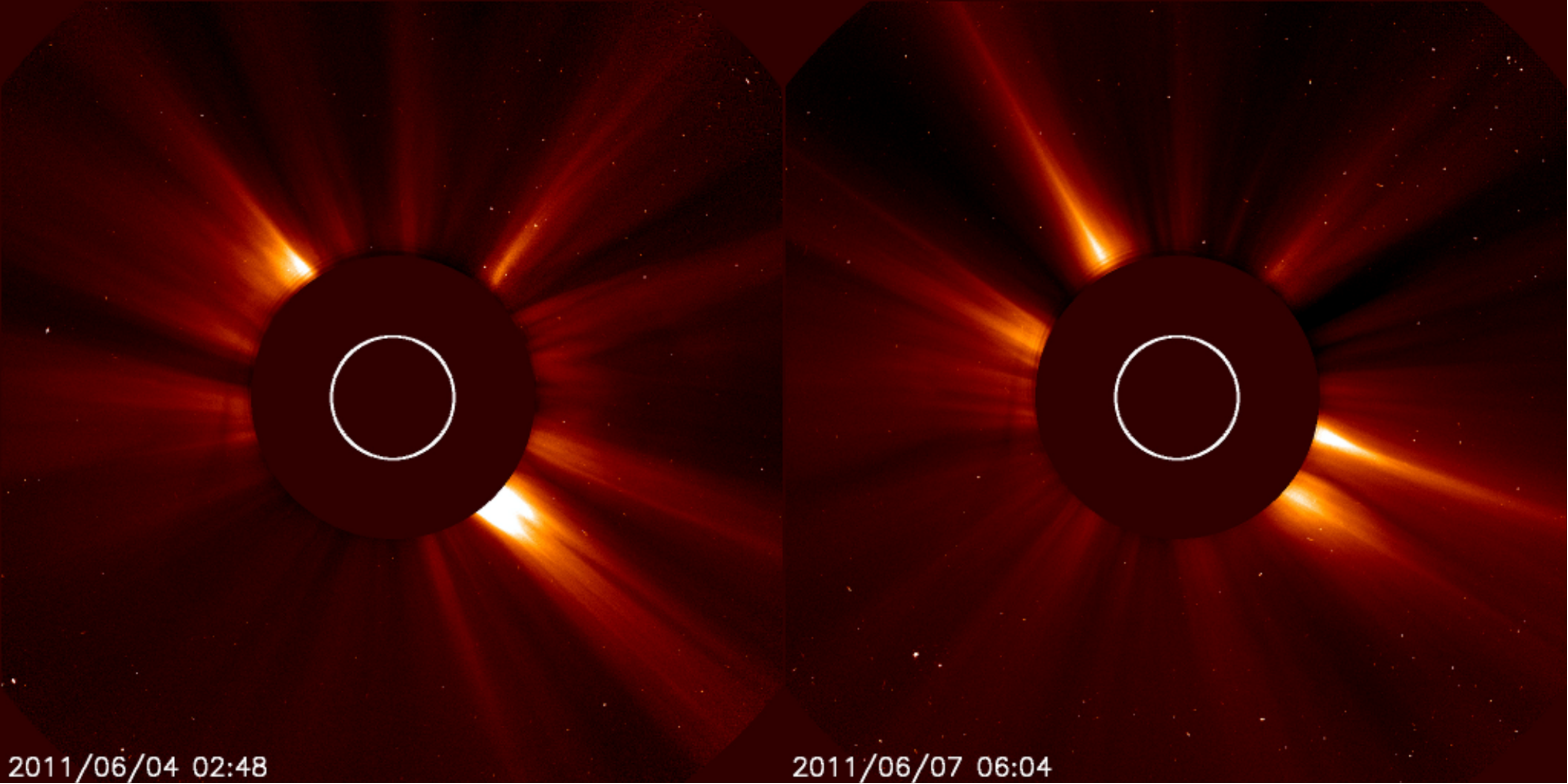}
\caption{Appearance of the white light corona as observed on June 4, 02:48 UT (left) before the acquisition of the pB image used for the coronal density determination, and on June 7, 06:04 UT (right) before the occurrence of the eruption.}\label{fig:corona}
\end{figure*}

\subsubsection{Pre-CME coronal densities}
For the density calculation we use SOHO/LASCO C2 polarized brightness (pB) images. It is well known that the K-corona brightness originates from Thomson scattering of photospheric light by free electrons in the solar corona \citep[e.g.,][]{bil66}. Because the emission is optically thin, the observer sees a contribution from electrons located all along the line of sight. In addition to the K-corona, observations will contain a component due to scattering of photospheric light from interplanetary dust (the so-called F-corona). This component must be eliminated from the data to derive the coronal electron density; however, in the case of pB observations at small altitudes ($\lesssim 5$~\rsun), the F corona can be assumed unpolarized and thus does not contribute to the pB \citep[][]{hay01}.

The intensity of the scattered light depends on the number of scattering electrons and several geometric factors, as was first outlined by \citet[][]{min30}. In the absence of F corona, the polarized brightness observed on the plane of the sky is given by the following equation:
\begin{equation}\label{eq:vandehulst}
\text{pB}(\varrho)=C\int_\varrho^\infty n_e(r)\left[A(r)-B(r)\right]\frac{\varrho^2\,dr}{r\sqrt{r^2-\varrho^2}},
\end{equation}
where $C$ is a unit conversion factor, $n_e$ is the electron density, $A$ and $B$ are geometric factors \citep[][]{vdh50,bil66}, $\varrho$ is the projected heliocentric distance of the point (impact distance), and $r$ is the actual heliocentric distance from Sun center. The integration is performed along the line of sight through the considered point. \citet[][]{vdh50} developed a well known method for estimating the electron density by the inversion of Equation (\ref{eq:vandehulst}) under the assumptions that: (1) the observed polarized brightness along a single radial can be expressed in the polynomial form $\text{pB}(r)=\sum_k \alpha_k r^{-k}$ and (2) that the coronal electron density is axisymmetric. We apply this method to the latest LASCO C2 pB image acquired before the June 7th CME, in order to determine the pre-CME electron density distribution in the corona.

The pB image considered here is obtained from the polarization sequence of observations recorded on June~4th 2011, starting at 02:54~UT, i.e. about three days before the occurrence of the June 7 CME. During this three-day time lag, three other much smaller CMEs occurred having a central propagation direction in the same latitudinal sector crossed by the June 7th CME ($70^\circ$S--$40^\circ$N), as reported in the SOHO/LASCO CME catalog: on June~4th, at 06:48~UT and 22:05~UT, and on June~6th, at 07:30~UT. Nevertheless, despite these smaller scale events and coronal evolution, a direct comparison between the LASCO C2 white-light images acquired on June~4th at 02:48~UT and on June~7th immediately before the eruption at 06:04~UT shows that the overall density structure of the corona above the west limb of the Sun is quite similar even after more than three days (Figure \ref{fig:corona}), hence the electron density estimated from the inversion of the June~4th pB data can be considered at least a first order approximation of the real pre-CME coronal density configuration.

\begin{figure*}
\centering
\subfigure[]{\includegraphics[width=0.27\textwidth]{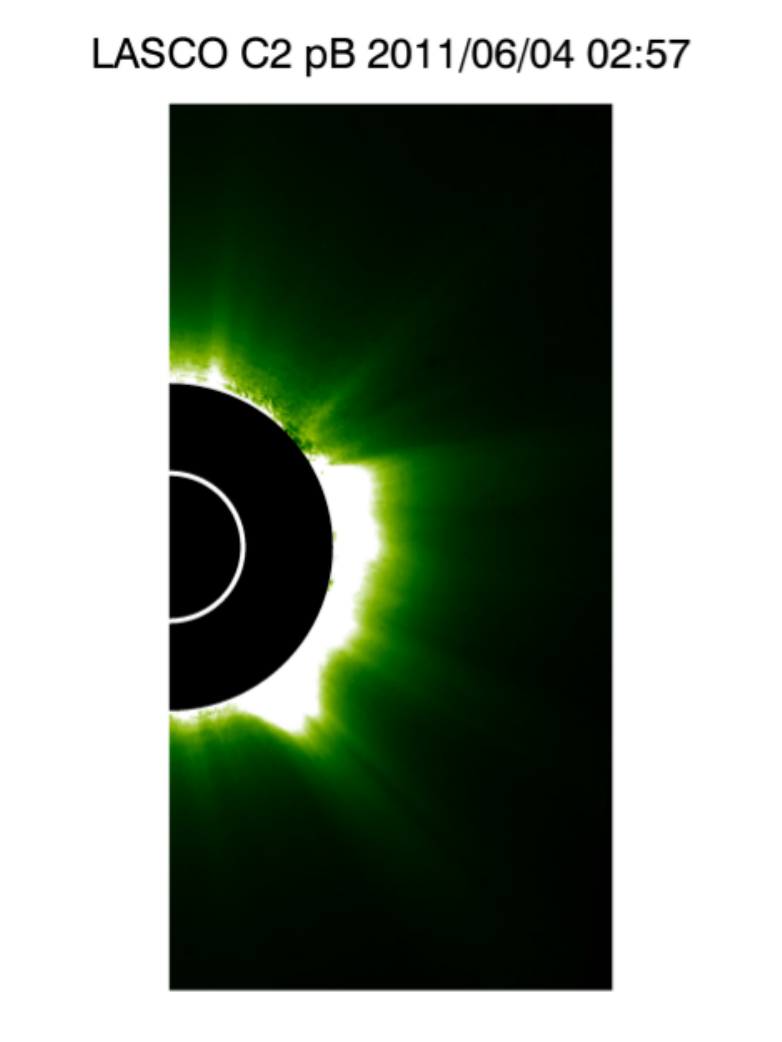}}
\subfigure[]{\includegraphics[width=0.72\textwidth]{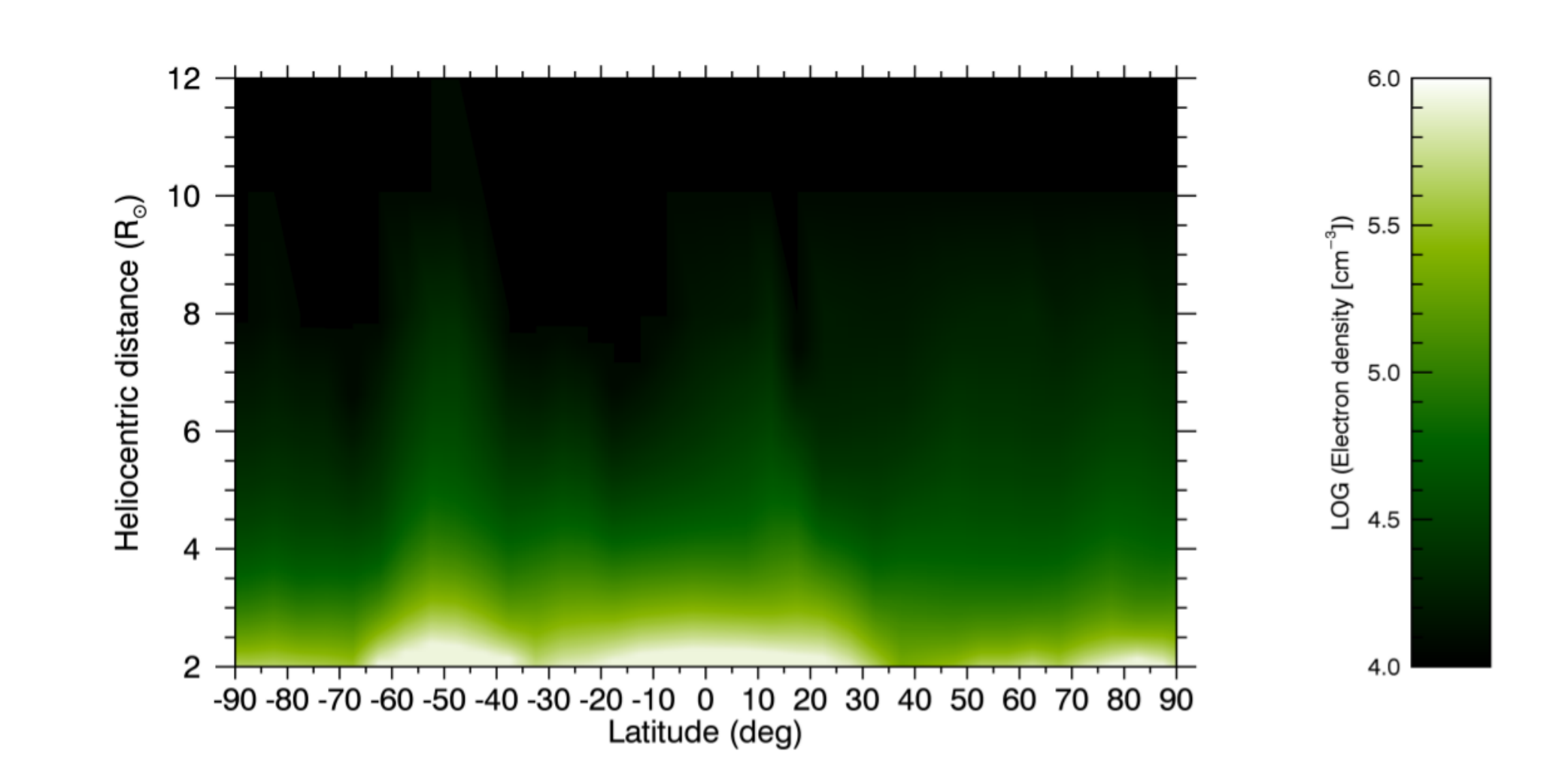}}
\caption{LASCO C2 polarized-brightness image of the solar corona above the west limb, acquired on June 4th 2011 at 02:57~UT (a) and the corresponding 2D electron density map derived from the inversion of the pB data (b).}\label{fig:density}
\end{figure*}

The electron density radial profiles obtained at different latitudes from the pB image (Fig.~\ref{fig:density}a) are combined into a 2D map in polar coordinates, shown in Fig.~\ref{fig:density}b. The map shows the density distribution in the latitudinal region being crossed later on by the shock, for heliocentric distances ranging between 2 and 12~\rsun; electron densities at distances from the Sun larger than 6~\rsun\ (the outer limit of the LASCO C2 field of view) are obtained through a power-law extrapolation of the density profiles assuming a radial dependence proportional to $r^{-2}$. The presence of the coronal streamer centered around $50^\circ$S, that is persistent till June~7, is very clear as it is associated with a local electron density maximum. Notice that in general coronal features are much less evident in the pB image and in the density map (Figure \ref{fig:density}) with respect to the regular LASCO frames (Figure \ref{fig:corona}) because the latter are obtained after subtraction of a monthly minimum background average to enhance the visibility of fainter structures.

\subsubsection{Shock position and kinematics}

White-light coronagraphic images can be used to identify the shock front location at different times and to distinguish between the shock-compressed plasma and the CME material, as extensively demonstrated by several works \citep[e.g.,][]{vourlidas03,ont09,bem10,bem11}. The CME-driven shock front can be identified as a weak brightness increase located above the expanding CME front, that is generally interpreted as the visible signature of the downstream plasma compression and density enhancement caused by the transit of the shock; for this reason, the shock front becomes visible only when the intensity scale of WL images is adjusted to bring out the fainter structures.

In this work, we determine the location of the shock front in both LASCO C2 and C3 total brightness images using a common procedure that consists of three steps: (1) we compute excess-mass (or base-difference) images by subtracting from each calibrated LASCO frame an average pre-event image that is representative of the quiescent background corona \citep[see][]{vou00,ont09}; (2) we apply a Normalizing Radial Graded Filter (NRGF), as described by Morgan et al. (2006), in order to reveal faint emission features at high heliocentric distances in the corona (this is particularly useful for the identification of the shock front in LASCO C3 images); (3) we measure the projected altitude of the shock by locating the intensity jump at the front in the radial direction. With this technique the location of the shock can be identified with an estimated uncertainty of $\pm 3$ pixels on average and $\pm 5$ pixels for LASCO C2 and C3 images, respectively. Larger uncertainties could be related with the applied procedure of background subtraction, in the possible locations where the pre-eruption corona significantly changed during the event.

\begin{figure}
\centering
\includegraphics[width=\columnwidth]{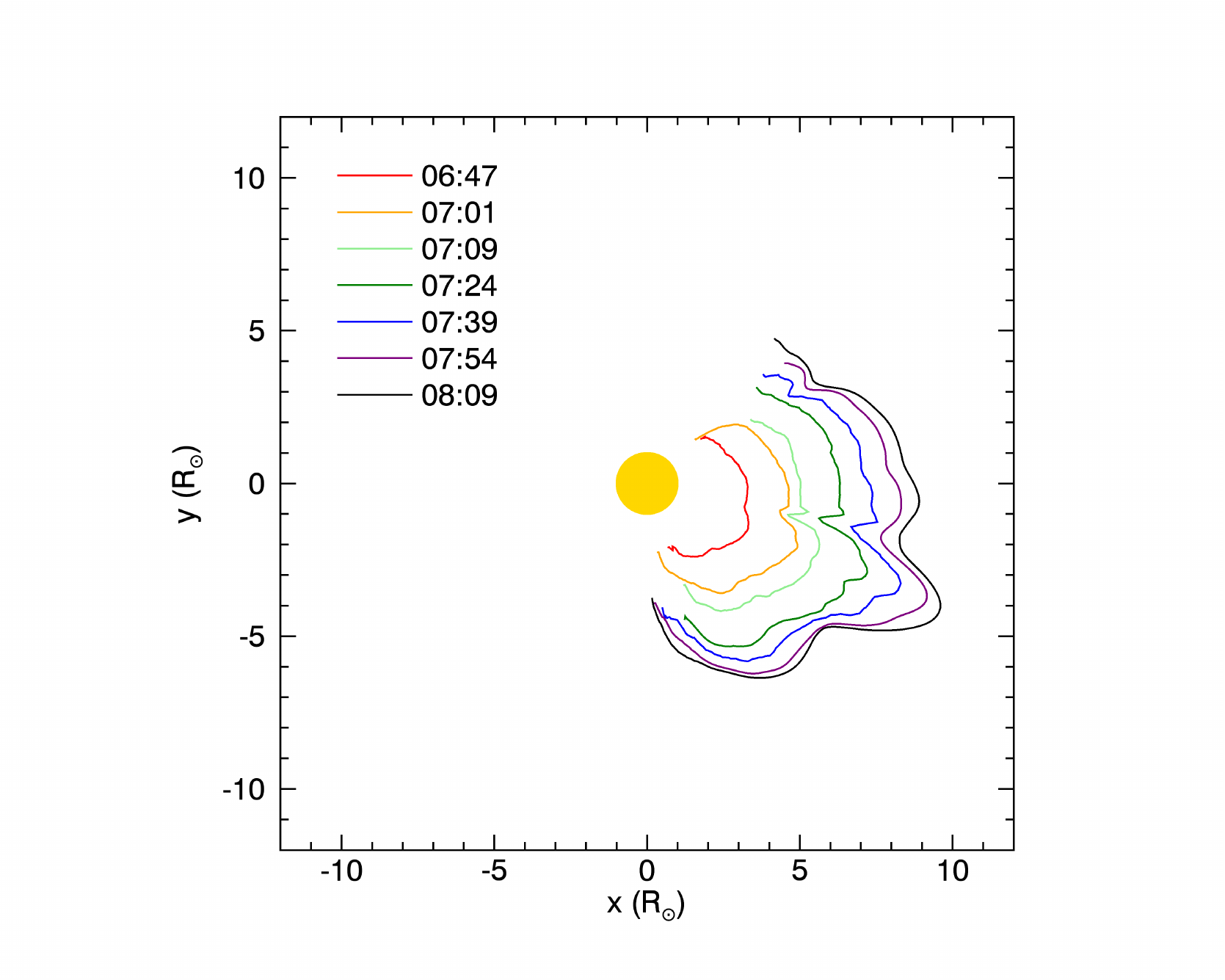}
\caption{Cartesian plot showing the locations of the shock front identified at different times in LASCO C2 and C3 white-light images.}\label{fig:positions}
\end{figure}

\begin{figure}
\centering
\includegraphics[width=\columnwidth]{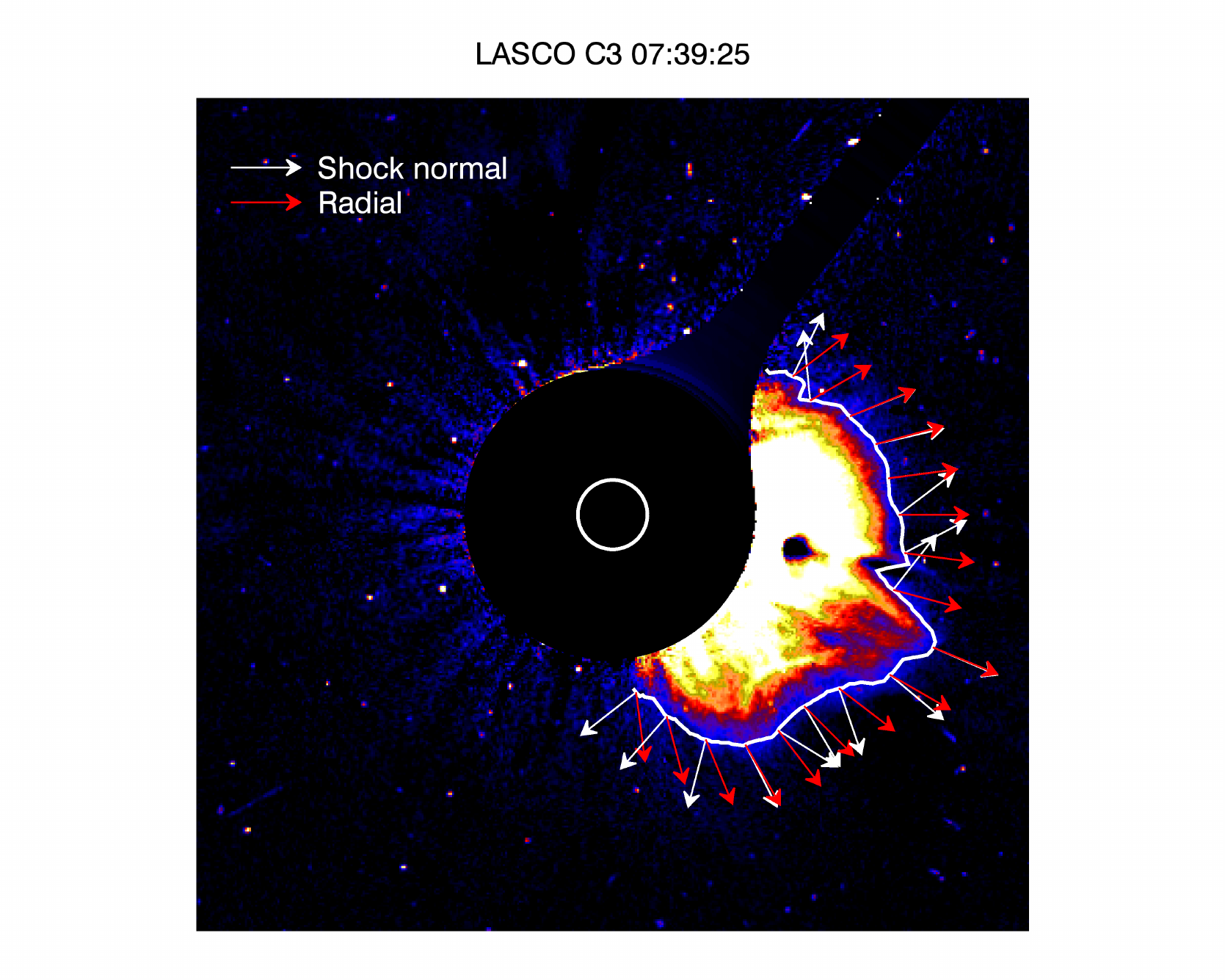}
\caption{Base-difference LASCO C3 image showing the location of the shock front (solid white line) at 07:39~UT and a schematic representation of selected vectors normal to the shock surface (white arrows) and corresponding radial directions in the same points (red arrows).}\label{fig:lasco_norm}
\end{figure}

We apply this procedure to seven consecutive images where we could identify signatures of the shock: two from LASCO C2, acquired at 06:47 and 07:01~UT, and five from LASCO C3, acquired at 07:09, 07:24, 07:39, 07:54, and 08:09~UT, respectively (see Figure \ref{fig:event}). Later on, we were not able to locate the shock front with a significant accuracy in LASCO C3 images. The curves giving the position the shock fronts identified in the considered WL images are plotted in Figure~\ref{fig:positions}. The shock appears to propagate almost symmetrically and to exhibit only a moderate latitudinal displacement, since the center of the shock (i.e., the highest point along the front) has a latitudinal location which is always in the range 21--25$^\circ$S. We notice here that around a latitude of about 12$^\circ$S the identified location of the shock surface shows a clear discontinuity, which is likely due to the Northward displacement of a the pre-event coronal streamer, leading to an overestimate (underestimate) of the shock projected altitude Northward (Southward) of the streamer itself.

These curves can be easily employed to derive, all along each shock front, the angle $\theta_\text{sh}$ between the normal to the shock front and the radial direction, as well as the latitudinal distribution of the average shock speed, $v_\text{sh}$. These quantities are essential for the determination of the Alfv\'enic Mach number and the upstream plasma velocity distribution, as discussed in the following section. As an example, Figure~\ref{fig:lasco_norm} shows the relative orientation of vectors parallel with the radial direction and those normal to the shock surface at different positions along the front as we identified in the LASCO C3 image acquired at 07:39~UT. It is evident from this Figure that $\theta_\text{sh}$ angles are in general larger at the flanks of the shock, and smaller near the shock center (or ``nose''). This result confirms what we already found in recent works \citep[see, e.g.,][]{bem14} and suggests that we may expect the prevalence of quasi-perpendicular shock conditions at the flanks and quasi-parallel shock conditions at the center of the shock.

\begin{figure*}
\centering
\vspace{-4cm}
\includegraphics[width=\textwidth]{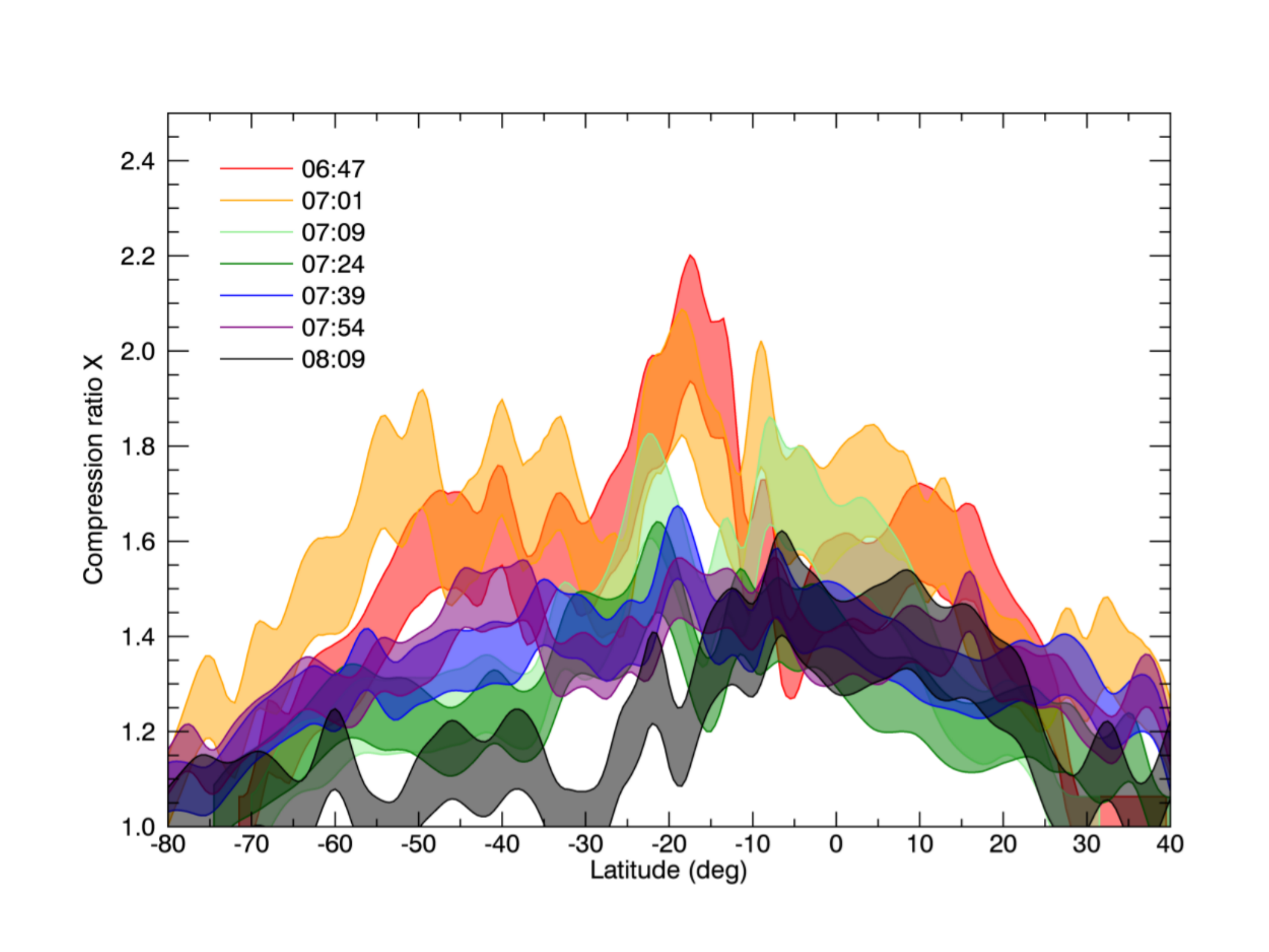}
\caption{Compression ratios $X \equiv \rho_\text{d}/\rho_\text{u}$ as measured along the shock fronts identified in LASCO observations and reported in Fig.~\ref{fig:positions}. Each profile is shown as a thick shaded area representing the uncertainty in the derived $X$ values.}\label{fig:ratios}
\end{figure*}

The radial component of the average shock speed is obtained at each latitude simply as $v_\text{r}=\Delta\varrho/\Delta t$, where $\Delta\varrho$ is the variation of the projected heliocentric distance of the shock measured in the radial direction between two consecutive shock curves. The true shock velocity can be then derived simply as $v_\text{sh}=v_\text{r}\cdot\cos\theta_\text{sh}$. Note that, as in \citet[][]{bem14}, this corresponds to assume isotropic self-similar expansion of the front in the range of common latitudes between consecutive curves, but taking into account the correction for the latitudinal shock propagation. A 2D polar map of radial velocity distribution $v_\text{r}$ in the region where the shock propagates is obtained by interpolating with polynomial fitting the heliocentric distance values at each latitude and altitude along the shock fronts, and is shown in Figure~\ref{fig:results1} (top-left panel). The resulting radial shock speed is (as expected) larger at the center of the shock at all altitudes, then it decreases toward the shock flanks; at a heliocentric distance of $2.5$~\rsun\ it reaches a value as high as $\sim 1200$~km~s$^{-1}$ near the center and $\sim 800-900$~km~s$^{-1}$ $\sim 20^\circ$ away from it. The shock also appears to decelerate during its propagation, since the velocity at higher altitudes is progressively smaller: for instance, at 12~\rsun\ $v_\text{sh}\simeq 1000$~km~s$^{-1}$ at the shock center. This means that the shock is losing its energy as it expands; this is also supported by the results we obtain for the compression ratio and the Alfv\'enic Mach number, as discussed in the following section.

\subsubsection{Compression ratio, Alfv\'enic Mach number, and Alfv\'en speed}

The shock compression ratio $X$, defined as the ratio between the downstream (i.e., post-shock) and the upstream (i.e., pre-shock) plasma densities, $X\equiv \rho_\text{d}/\rho_\text{u}$, is determined here as described in \citet[][]{bem11}. For each pixel along an identified shock front, we measure the total white-light brightness of the compressed downstream plasma, tB$_\text{d}$, from the corresponding LASCO C2 or C3 image, and, at the same locations in the corona, the upstream brightness tB$_\text{u}$ from the last image acquired before the arrival of the shock. This provides us with the observed ratio $(\text{tB}_\text{d}/\text{tB}_\text{u})_\text{obs}$.

On the other hand, the upstream total brightness tB$_\text{u}(\varrho)$ expected at a projected altitude $\varrho$ in the corona can be evaluated through the line-of-sight integration of the upstream electron density profile, $n_e(r)$, multiplied by a geometrical factor $K$ that includes all the geometrical parameters for Thomson scattering:
\begin{equation}
\text{tB}_\text{u}(\varrho)=\int_{\varrho}^{\infty}{K(r,\varrho)\cdot n_e(r)\,dr},
\end{equation}
where $r$ is the heliocentric distance of the scattering point along the line of sight. The expected downstream total brightness tB$_\text{d}$ is similarly given by the sum of two integrals: one performed over the unshocked corona (with density $n_e$) and the other over a length $L$ across the shocked plasma with density $X\cdot n_e$ ($X \geq 1$):
\begin{eqnarray}
&&\text{tB}_\text{d}(\varrho)=\int_{\varrho}^{\infty}{K(r,\varrho)\cdot n_e(r)\,dr} +\\ \nonumber 
&&\int_{\varrho}^{r_\text{sh}}{K(r,\varrho)\cdot (X-1)\cdot n_e(r)\,dr},
\end{eqnarray}
where $r_\text{sh}=\sqrt{\varrho^2+L^2}$ and $X$ is precisely the unknown compression ratio. The shock depth $L$ is estimated as in \citet[][]{bem10}, i.e., by assuming that the shock surface has the three-dimensional shape of an hemispherical shell with thickness equal to the 2D projected thickness $d$ of the white-light intensity jump across the shock, corrected for the shock motion during the LASCO C2 or C3 exposure time. For each frame we estimated an average value of the shock depth $L$, and applied the same value to the whole shock front. Given $L$ and by adopting the radial density profiles derived from the analysis of the LASCO C2 pB, the shock compression ratio $X$ can be inferred directly from the comparison between the observed and the expected total brightness ratios: $(\text{tB}_\text{d}/\text{tB}_\text{u})_\text{obs}=(\text{tB}_\text{d}/\text{tB}_\text{u})_\text{exp}$.

The corresponding curves for the compression ratio $X$ measured along the shock fronts with different LASCO C2 and C3 frames are reported in Figure~\ref{fig:ratios}. The uncertainties in $X$ values shown in this Figure are due to the uncertainty in the identification of the exact location of the shock in C2 and C3 images (see above). The compression ratio reaches the maximum value of $\sim 2.1$ at 06:47~UT in a point that is very close to center of the shock front at that time located around a latitude of -20$^\circ$S; this $X$ value is quite lower than the upper limit adiabatic compression of 4 expected for a monoatomic gas. In all cases, the latitudinal dependence is similar: $X$ has a maximum around the center of the shock front, progressively but not monotonically decreasing toward the flanks. As the shock expands, the $X$ values decrease on average all along the shock fronts: for instance, at 08:09~UT the maximum value is of $\sim 1.5$; as already pointed out in the previous section, this indicates that the shock is dissipating its energy while propagating in the corona. These results are in agreement with those reported by \citet[][]{bem11} in their analysis of a different CME-driven shock. We notice here that, as explained above, the $X$ values have been not derived after background subtraction, but from the ratio between the total brightnesses observed at the shock location and those observed at the same pixels in the frame acquired just before the arrival of the shock. This method allows to remove in the ratio any possible uncertainty due to the instrumental calibration; moreover, because the shock is the faster feature propagating outward, no significant changes occurred in the corona aligned along the LOS between the two frames other than the compression due to the shock.

\begin{figure*}
\centering
\vspace{-3cm}
\subfigure[]{\includegraphics[width=0.49\textwidth]{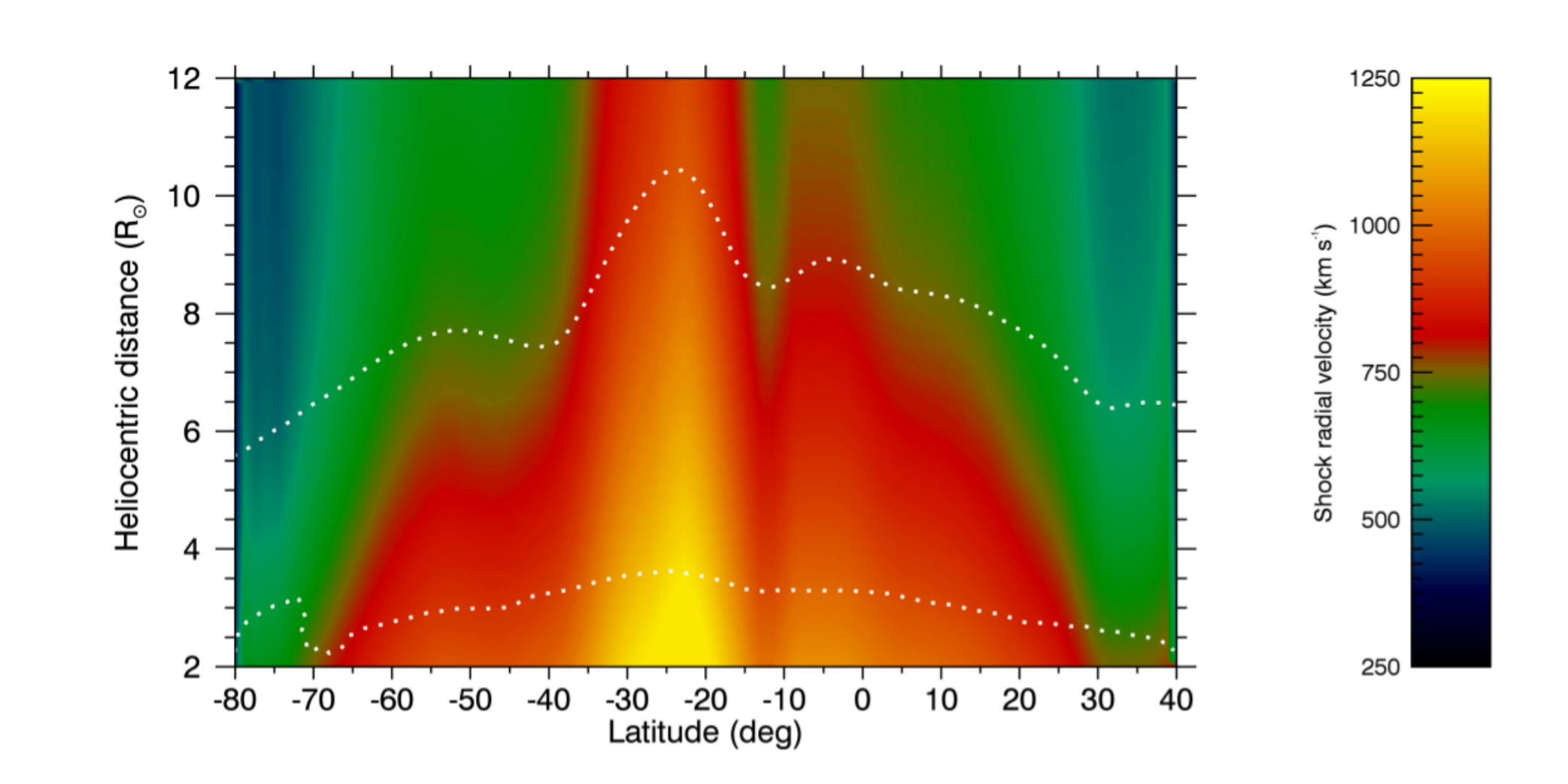}}
\subfigure[]{\includegraphics[width=0.49\textwidth]{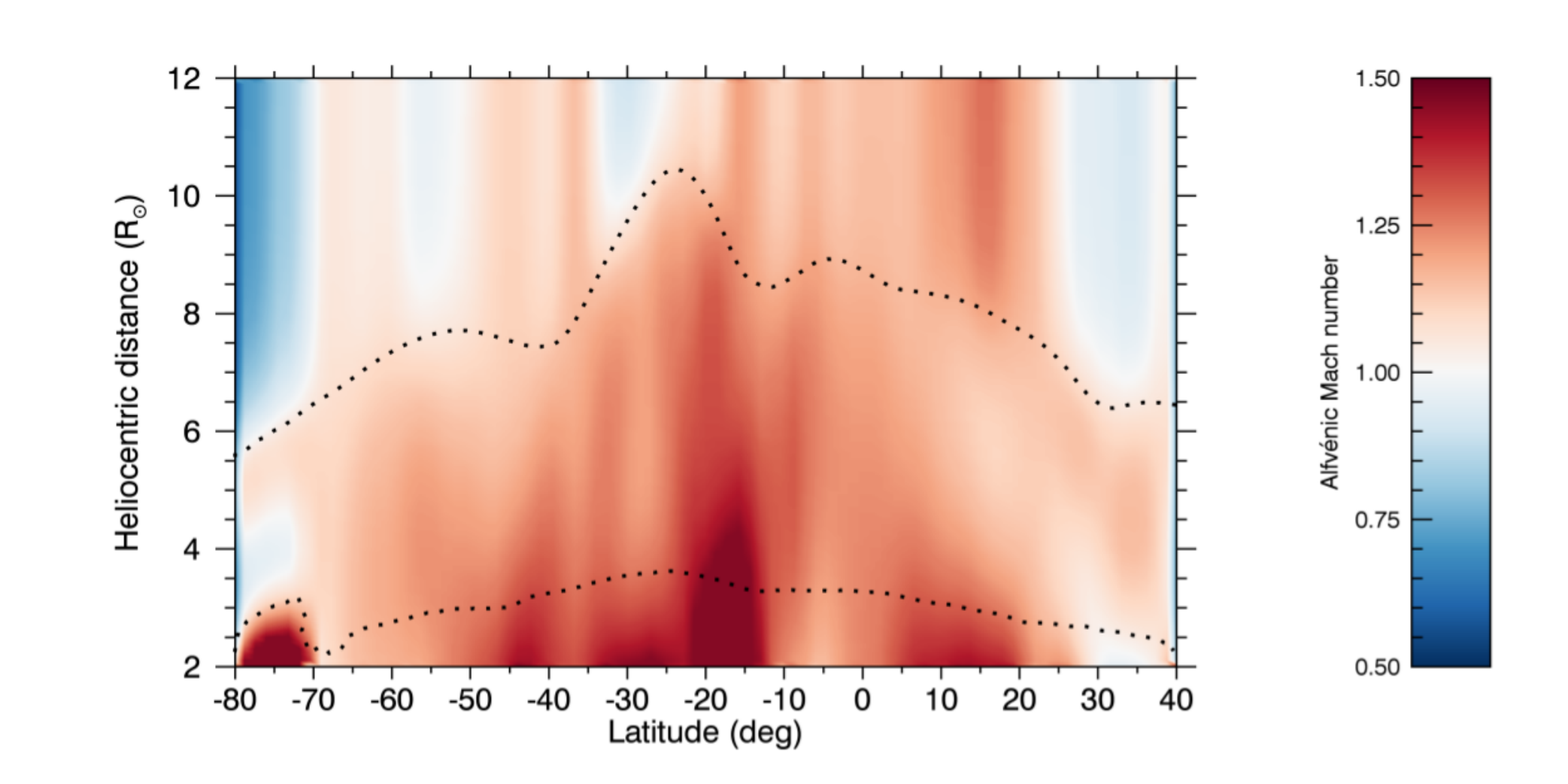}}
\subfigure[]{\includegraphics[width=0.49\textwidth]{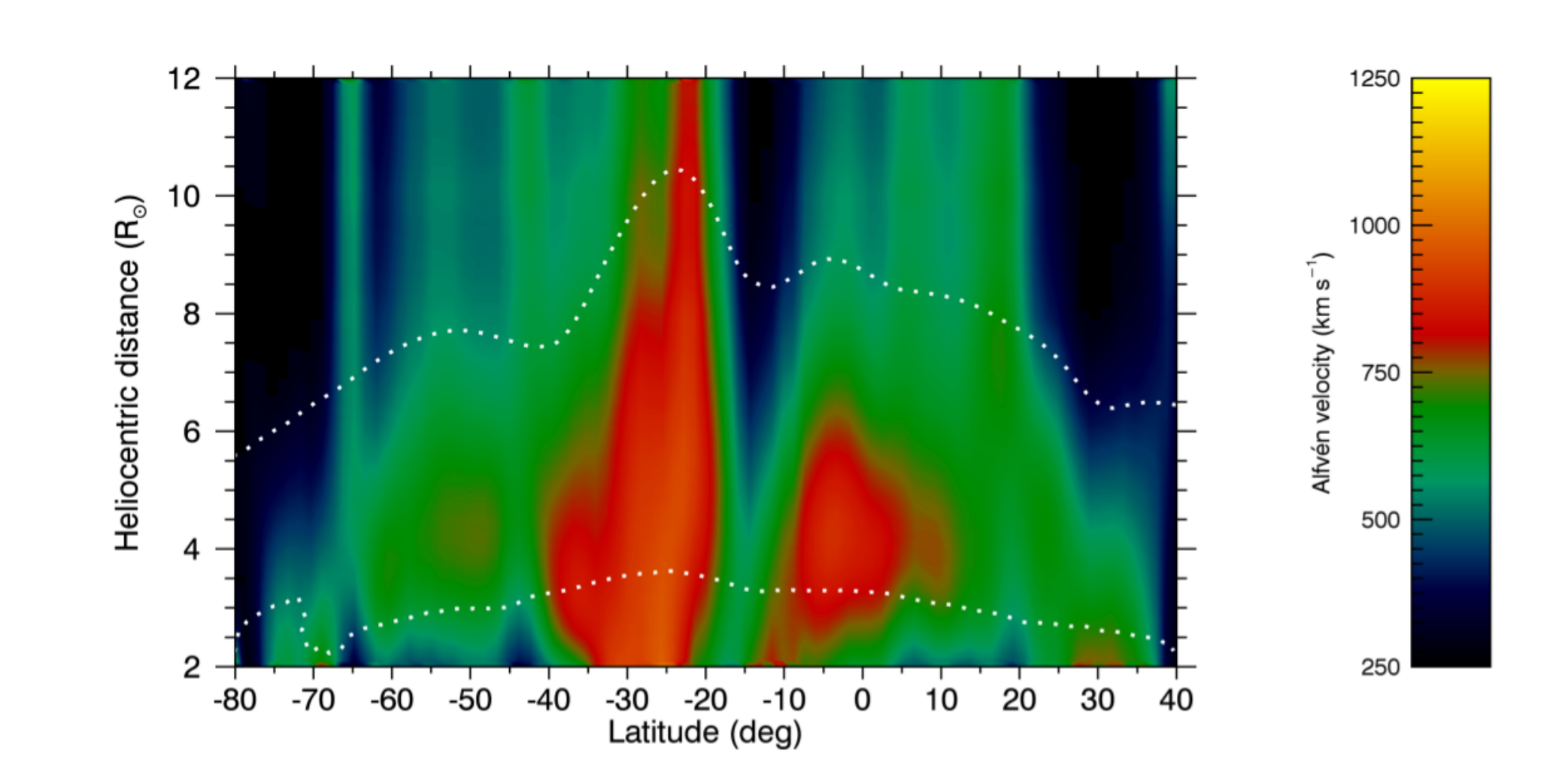}}
\subfigure[]{\includegraphics[width=0.49\textwidth]{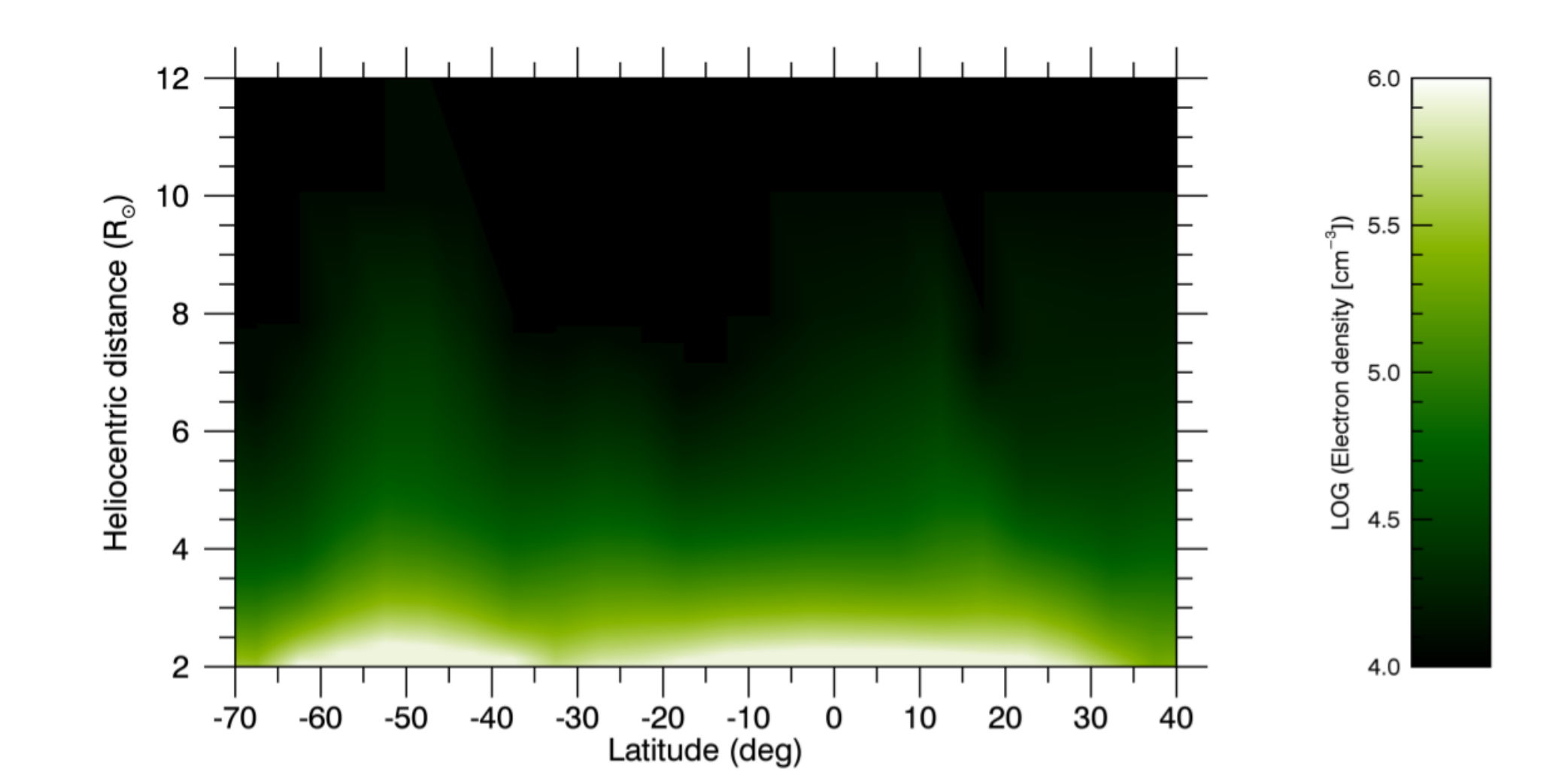}}
\caption{2D maps showing the distribution of the radial shock velocity $v_\text{r}$ (a), the Alfv\'enic Mach number $M_\text{A}$ (b), the Alfv\'en speed $v_\text{A}$ (c), and as a reference the pre-shock coronal densities $n_e$ (d). The $M_\text{A}$ and $v_\text{A}$ values are derived by assuming a negligible solar wind speed, as described in the text. In each panel real measurements were obtained only in the region between the two dotted lines, while values shown out of these region have been extrapolated at higher and lower altitudes.}\label{fig:results1}
\end{figure*}

The Alfv\'enic Mach number is defined as the ratio between the upstream plasma velocity $v_\text{u}$ (i.e., the velocity of the plasma flowing toward the shock surface in the reference frame at rest with the shock itself) and the Alfv\'en speed $v_\text{A}$, $M_\text{A} \equiv v_\text{u}/v_\text{A}$. $M_\text{A}$ can be estimated from the compression ratio $X$ and the angle $\theta_\text{sh}$ under two assumptions: (1) the plasma $\beta \ll 1$ ($\beta$ is the ratio between the thermal and magnetic plasma pressures) and (2) the upstream magnetic field is radially directed, so that the angle between the shock normal and the magnetic field vector can be assumed to be equal to $\theta_\text{sh}$ on the plane of the sky. These are not strong assumptions, as discussed in \citet[][]{bem11}, and can be considered fairly verified also in our case. Under these hypotheses, as we verified observationally in \citet[][]{bem14} and theoretically in \citet[][]{bacchini2015}, the Alfv\'enic Mach number is well approximated in the general case of oblique shock by the following semi-empirical formula:
\begin{equation}\label{eq:mach}
M_{\text{A}\angle}=\sqrt{M_{\text{A}\parallel}^2\cos^2\theta_\text{sh}+M_{\text{A}\perp}^2\sin^2\theta_\text{sh}},
\end{equation}
where $M_{\text{A}\parallel}=\sqrt{X}$ and $M_{\text{A}\perp}=\sqrt{\frac{1}{2}X(X+5)/(4-X)}$ are the expected Mach numbers for parallel and perpendicular shocks, respectively, for a $\beta \ll 1$ plasma. The validity of Eq.~(\ref{eq:mach}) has been confirmed by the analysis of \citet[][]{bem14} which takes advantage of both white-light and ultraviolet data from the \emph{Ultra-Violet Coronagraph Spectrometer} (UVCS) on board SOHO (see discussion therein) and has been recently tested with MHD numerical simulations by \citet[][]{bacchini2015}. This equation allowed us to derive, from different values of $X$ and $\theta_\text{sh}$ parameters, 2D polar maps of $M_{\text{A}\angle}$ values, as shown in Figure~\ref{fig:results1} (top right panel). This map clearly shows that in the early phases the shock was super-Alfv\'enic at all latitudes (with larger $M_\text{A}$ values at the shock nose), while later on (i.e. higher up) keeps super-Alfv\'enic numbers only at the nose.

The Alfv\'en speed can be derived, in turn, from $M_\text{A}$ values once the upstream plasma velocity is known or estimated. The upstream velocity is given by $v_\text{u}=|\mathbf{v_{sw}}-\mathbf{v_{sh}}|$, where $\mathbf{v_{sw}}$ is the outflow solar wind speed, assumed to be radial, and $\mathbf{v_{sh}}$ is the shock speed. In our case, we have no direct measurements of the wind flows in the corona, hence we must adopt a model for the solar wind expansion in order to infer the Alfv\'en speed from the Alfv\'enic Mach number. To this end, a first-order approximation can be obtained by assuming $\mathbf{v_{sw}}=\mathbf{0}$ in the previous equation, i.e., by neglecting the solar wind at all. This is not a realistic assumption, but it is rather reasonable, considering that at low altitudes in the corona ($\lesssim 5$~\rsun) and in the early phase of propagation, the shock speed may be up to one order of magnitude larger than typical wind velocities measured outside coronal holes \citep[$\approx 100$--300~km~s$^{-1}$; see, e.g.,][]{sus08}. Under this hypothesis, the estimated Alfv\'en speed can be considered as an upper limit to the real values. Possible consequences of this assumption will be discussed in the last Section.

\begin{figure*}
\centering
\vspace{-11.5cm}
\includegraphics[width=0.9\textwidth]{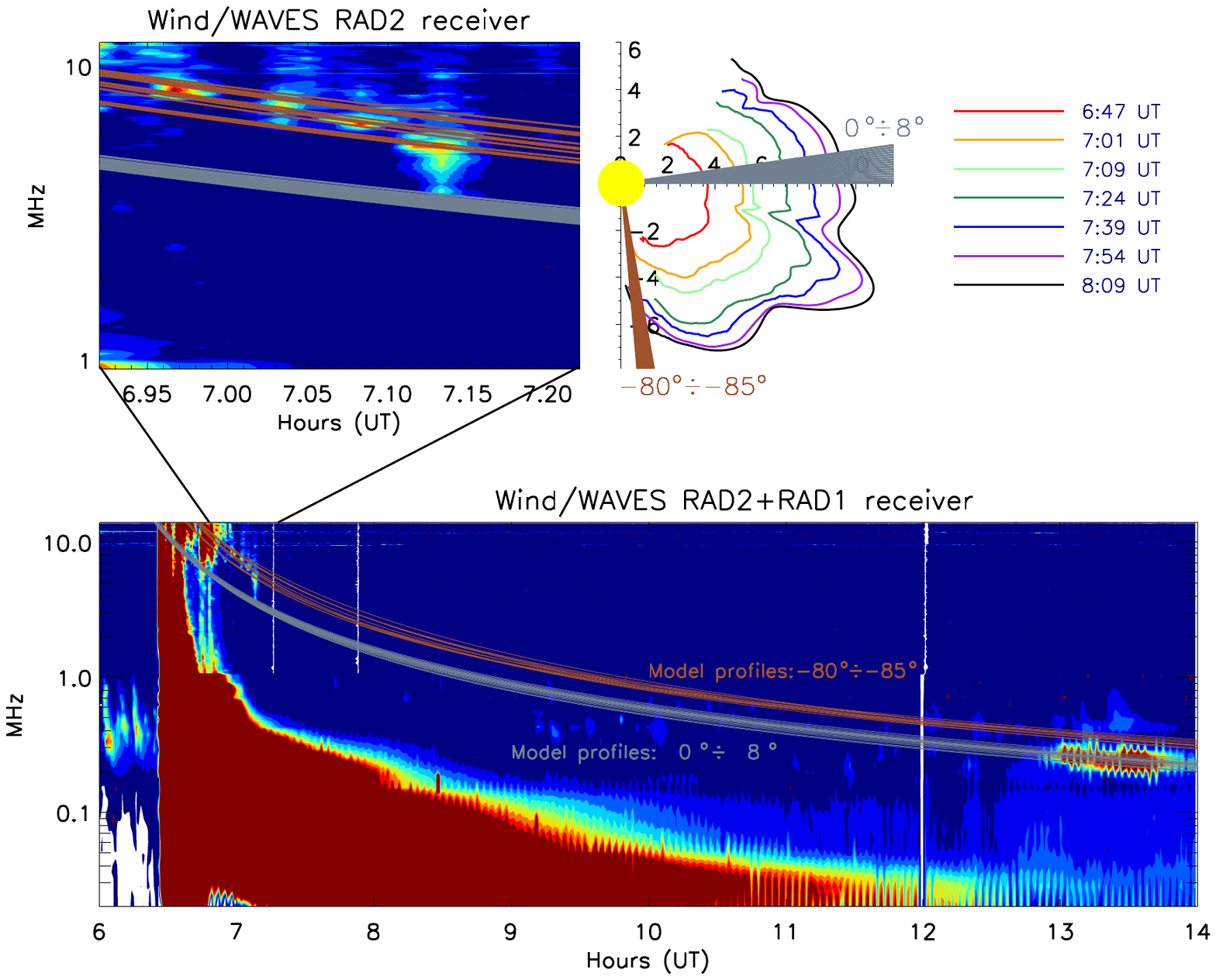}
\vspace{-2cm}
\caption{Lower panel: Dynamic spectrum of the Wind/WAVES radio data in the frequency range between 20 KHz and 13.8 MHz from 6 to 14 UT on 2011 June 7, showing, showing the decametric to kilometric type II radio emissions associated with the CME. The upper panel at the left shows details of the radio emission associated with the emission excited earlier at the southern flank of the shock. The curves on this plot are also explained in the text.}\label{fig:radio}
\end{figure*}

2D polar maps of the Alfv\'en speed are shown in Figure~\ref{fig:results1} (bottom-left panels); these maps have been obtained again with polynomial (third-order) interpolation of the Alfv\'en speeds measured at different locations (i.e. latitudes and altitudes) of the shock front at different times (Figure~\ref{fig:positions}). Results plotted in Figure ~\ref{fig:results1} clearly show that the Alfv\'en speed has not only radial, but also significant latitudinal modulations. The Alfv\'en speed reaches the highest value ($\sim 1000$~km~s$^{-1}$) at the lowest altitudes in the equatorial belt. The latitudinal dependence is rather complex, with an alternation of local minima and maxima ranging between $\sim 600$ and $\sim 1000$~km~s$^{-1}$. At increasing altitudes, $v_\text{A}$ generally decreases, with values that never exceed 800~km~s$^{-1}$ at 12~\rsun. Interestingly, the regions characterized by the slowest decrease in electron density (around $\sim 50^\circ$S and around $\sim 10^\circ$N; see Fig.~\ref{fig:density}) are also those where the Alfv\'en speed decreases more steeply, reaching values below $\sim 500$~km~s$^{-1}$ already at 5~\rsun. As a consequence, in the early propagation phase (i.e., at low altitudes) the shock is significantly super-Alfv\'enic not only at the nose but also in several regions distributed in the flanks of the shock surface. These high-density and high-Mach number regions are very probable candidates as sources of particle acceleration and type-II radio bursts; we discuss in the next section possible correlations with the sources of radio emission identified from radio dynamic spectra, while the determination of the magnetic field strength is discussed in the last Section.

\subsection{Radio dynamic specrum} \label{sec:radiodata}

As it is well known, shock waves are able to accelerate electron beams to suprathermal energies, which in turn can produce Langmuir waves that are converted by means of nonlinear wave-wave interactions into electromagnetic waves near the fundamental and/or harmonic of the local electron plasma frequency $f_{pe}$. Since the coronal density $n_e$ decreases with increasing heliocentric distance and $f_{pe} \propto {n_e^{1/2}}$, the expanding shock surface produces type-II radio emissions at decreasing frequencies as it propagates through space and the measured frequency drift rate at a given time is directly related to the shock speed. The observed frequency drift rate provides therefore information on the shock dynamics through the corona, while its onset depends on the local magnetosonic speed.

The dynamic spectrum in the lower panel of Figure \ref{fig:radio} shows the intensity of the radio data from 06:00 to 14:00~UT on 2011 June 7 in the frequency range between 20~KHz and 13.8~MHz measured by the RAD1 and RAD2 radio receivers of the WAVES experiment on the Wind spacecraft. A very intense complex type-III-like radio emissions was observed beginning at 6:24~UT. This fast-drifting radio emission can be interpreted as the first radio signature indicating the lift-off of the CME on the Sun \citep[e.g.,][]{reinerkaiser1999} and is probably originated by the reconfiguration of the magnetic field in the lower corona that allows the energetic electrons produced by the flare to escape into the interplanetary medium \citep{reiner2000}. Two slowly-drifting episodes of strong type-II emission were also observed in the decametric range around 07:00 UT (clearly visible in the expanded upper left panel of Figure \ref{fig:radio}) and after 09:00 UT, abruptly intensifying between 13:00 and 14:00~UT (lower panel of Figure \ref{fig:radio}). We interpret these bands of emissions, as usually assumed when only one band is visible, as second harmonics. The origin of the second harmonic emission in type-II bursts is well understood as a result of coalescence of two plasma waves into a transverse one at twice the plasma frequency. Less intense, additional slow-drifting, type-II-like radio emissions at different times and frequencies are also visible, probably originating from different portions of the super-Alfv\'enically expanding shock surface.

\begin{figure*}
\centering
\vspace{-2.5cm}
\includegraphics[width=0.9\textwidth]{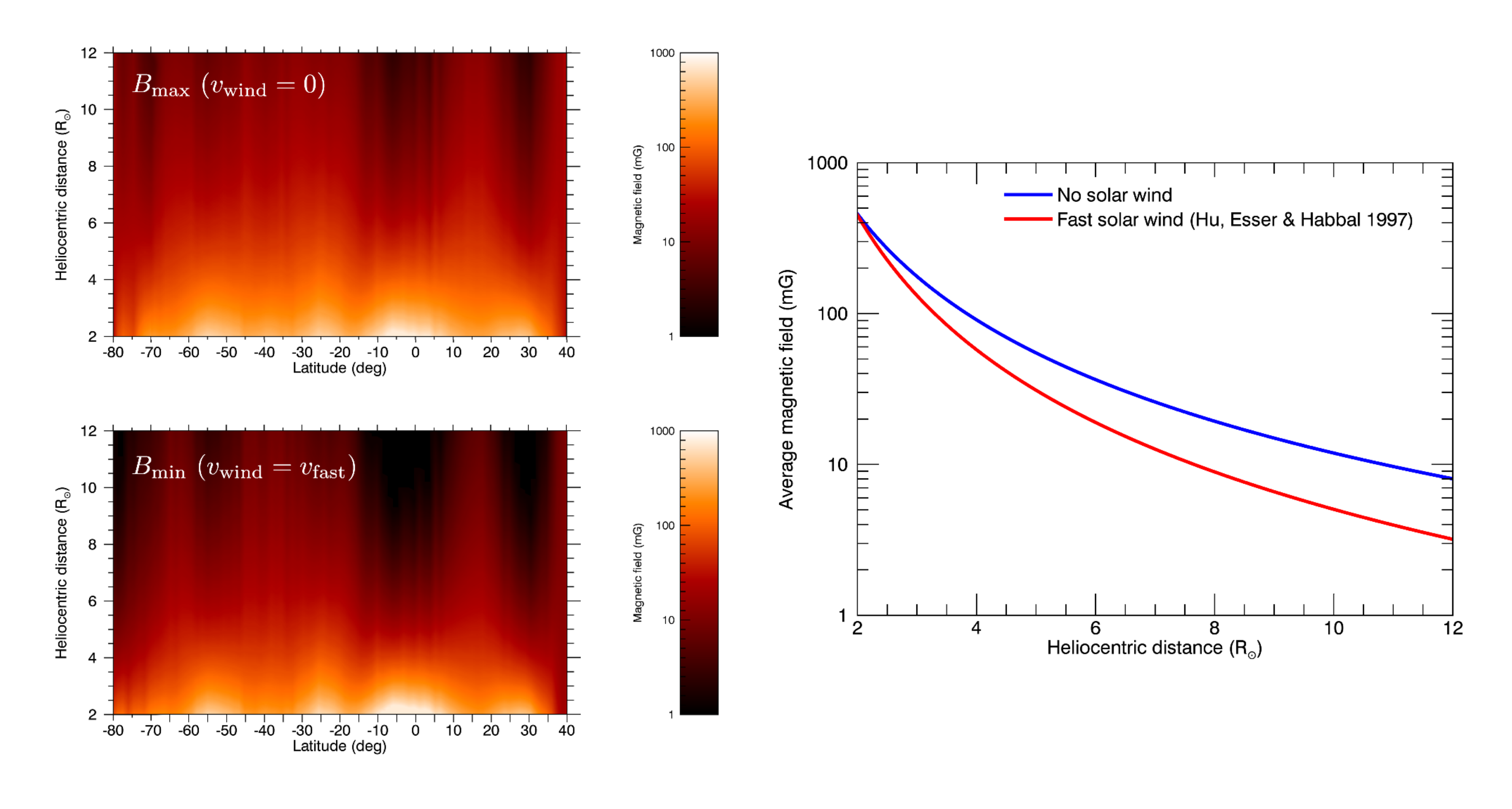}
\caption{Comparison between the 2D maps of coronal magnetic field strengths derived by assuming negligible wind speed (top left, upper limit for the field values) and fast wind speed at all latitudes (bottom left, lower limit for the field values). The right panel shows a comparison between the latitudinal average of magnetic fields obtained under the assumption of negligible wind speed (blue line) and assuming fast wind speed at all latitudes (red line).}\label{fig:fastslowwind}
\end{figure*}

In order to model the observed complex type-II radio emissions displayed in Figure \ref{fig:radio}, we need to know the coronal electron density profile at the time of the CME event. In fact, the density profile allows to convert the height measurements related to the shock surface dynamics to corresponding values of the coronal density as the frequencies $f$ are simply obtained as $f \approx f_{pe} \approx 9 \sqrt{n_e[\rm cm^{-3} ]}$~KHz. Instead of relying on a generic coronal electron density model, as usually done in the literature, we used the coronal electron density at different heliocentric distances and latitudes provided by the LASCO pB measurements discussed in the previous section. These density estimates, obtained for heliocentric distances greater than about 2~\rsun, correspond to radio frequencies below about 14~MHz, i.e., the range of radio emissions observed in the Wind/WAVES dynamic spectrum. By assuming, as usual, second harmonic type-II emission and using the coronal density distribution inferred from the available LASCO pB observations to relate the type-II frequencies to their heliocentric heights, we identified, knowing the shock's surface height from the previous analysis, a set of synthetic type-II profiles that were superimposed (as dashed lines in Figure \ref{fig:radio}) to the radio dynamic spectrum for comparison with the actual type-II emissions. This comparison allowed to characterize all observed type-II features and, in particular, two distinct regions (assuming radial propagation) along the shock's surface where the brightest radio emissions were most likely generated. An accurate estimate of the model radio profiles could only be obtained considering the coronal parameters outward from the flare longitude of 66$^\circ$ W and not from 90$^\circ$ W (plane of the sky). Unfortunately, at the time when the CME occurred, the STEREO-A and -B spacecraft were located at 94.9$^\circ$ and 93.0$^\circ$ from the Sun-Earth line, respectively. Hence, coronagraphic images acquired by the STEREO coronagraphs would not provide any useful information about the corona lying on the meridional plane at 66$^\circ$ W. Said that, although we assume that no significant temporal and longitudinal variations are present between the density profile we inferred on the plane of the sky and the density really met by the shock, this assumption is undoubtedly much more realistic with respect to the one that involves the adoption of a generic power-law density profile, as usually done in the literature for this kind of studies \citep[see e.g.][]{ra99, pojo2006, liu2009, kong2015, dorovskyy2015}.

\begin{figure}
\centering
\includegraphics[width=\columnwidth]{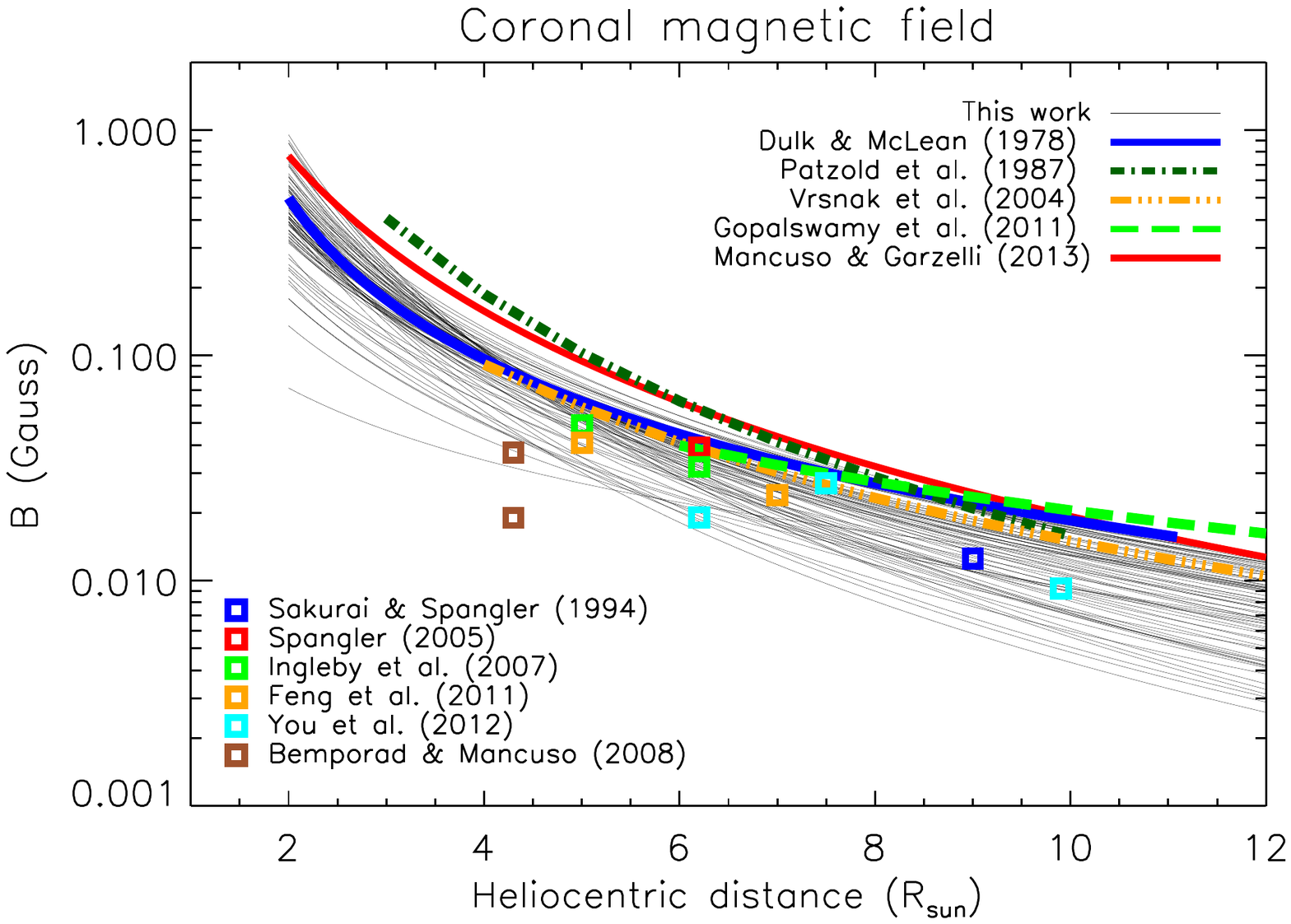}
\vspace*{-6cm}
\caption{Comparison between radial magnetic field profiles derived in this work at different latitudes (solid black lines), other magnetic field radial profiles provided in the literature (in particular: \citet{dulkmclean1978} - solid blue line, \citet{patzold1987} - dash-dotted dark green line, \citet{vrsnak2004} - dash-dotted orange line, \citet{gopalyashi2011} - dashed green line, and \citet{mancusogarzelli2013a} - solid red line) together with a compilation of other measurements (in particular: \citet{sakurai1994} - blue boxes, \citet{spangler2005} - red boxes, \citet{ingleby2007} - green boxes, \citet{feng2011} - orange boxes, \citet{you2012} - cyan boxes, and \citet{bem10} - brown boxes).}\label{fig:bmag_literature}
\end{figure}

With the above caveat in mind, we show that the two strong type-II bursts in this event are probably generated by two different portions of the shock (see upper right panel of Figure \ref{fig:radio}), one driven near the CME front and the other one at the southern flank region of the CME. We point out that the angular ranges specified in Figure \ref{fig:radio} are not intended to designate the accuracy of our results, but that they are simply meant to illustrate the angular location of the models that better fit the observed type II features. This result supports the scenario of type-II shock generation typically
arising at the CME flank due to interaction with a nearby streamer \citep[e.g.][]{ma04,ch08}. In this case, the type-II-emitting shock front may be quasi-perpendicular and thus apt to accelerate electrons by the shock drift acceleration mechanism \citep{ho83}.

\section{Discussion and Conclusions} \label{sec:concl}

The actual limitations in our understanding of many physical phenomena occuring in the solar corona is due in first place to our limited knowledge of the coronal magnetic fields. Knowledge of its strength and orientation is primarily based on extrapolations from observations of magnetic fields in the photosphere, where the magnetic field is strong and the Zeeman effect produces a detectable splitting of atomic levels and a subsequent polarization of the emitted light. Nevertheless, extrapolations from photospheric fields are model-dependent, static (no eruptive events) and fail to reproduce accurately complex coronal topologies. For these reasons, many different techniques have been developed to measure magnetic fields in the extended corona using radio observations and taking advantage of Faraday rotation \citep[e.g.,][]{maspan1999, mancusogarzelli2013a, mancusogarzelli2013b} and circular polarization in radio bursts \citep[e.g.,][]{hariharan2014}, or in the lower corona with EUV images using coronal seismology \citep[e.g.,][]{west2011} and field extrapolations bounded to 3D reconstructions \citep[e.g.,][]{aschwanden2014}. The recent development of spectro-polarimetric measurements of magnetic field strength and orientation is now providing very promising results \citep[e.g.,]{tomczyk2007, dove2011}, even if (due to the required polarimetric sensitivities) these techniques can be applied only in the lower corona ($h < 0.4$~\rsun).

\begin{figure*}
\centering
\vspace{-3cm}
\includegraphics[width=0.7\textwidth]{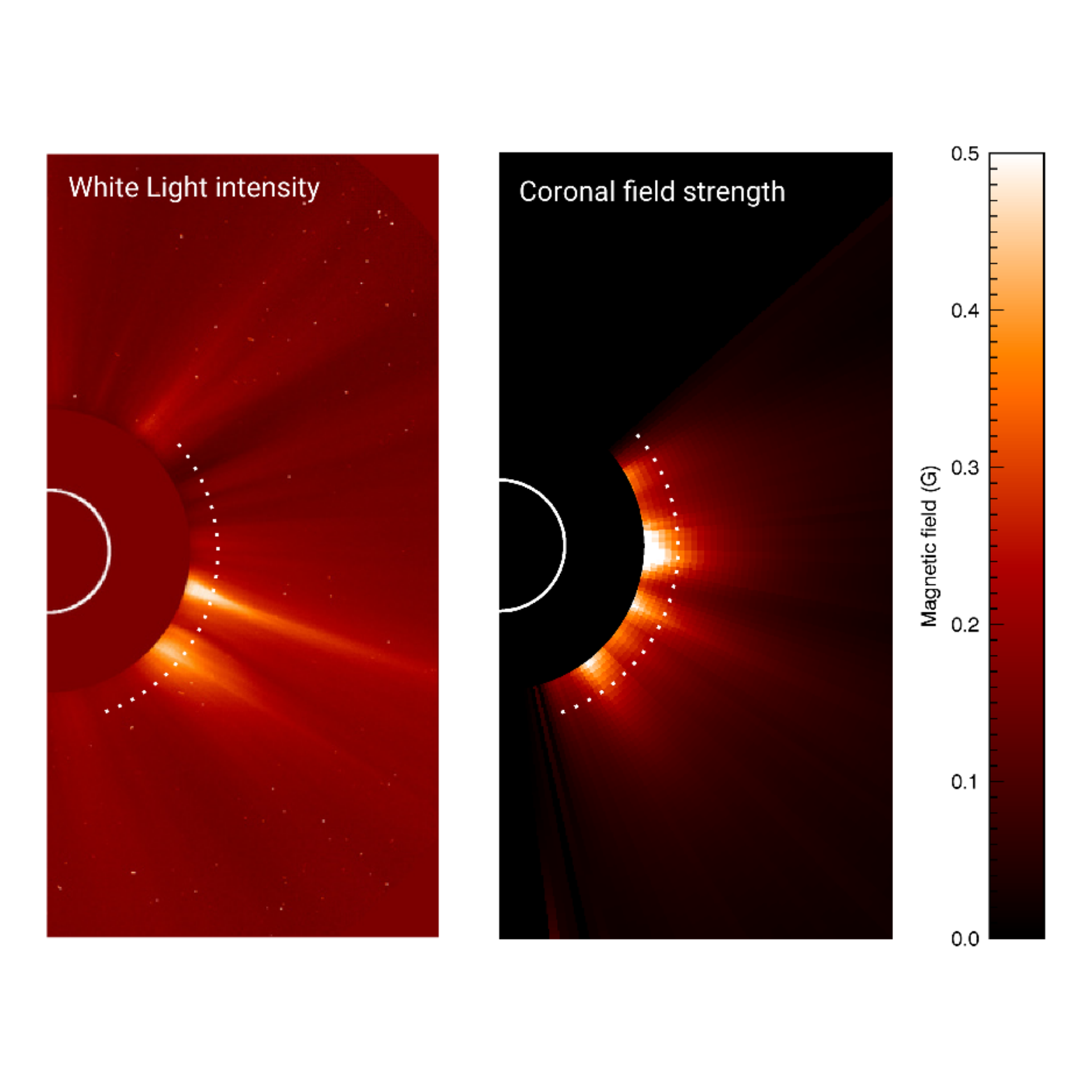}
\caption{Comparison between the pre-shock coronal white light structures observed by LASCO C2 coronagraph (left) and the magnetic field strengths derived in this work in the LASCO C2 field of view (right). The dashed lines show the location where latitudinal profiles of the WL intensity and field strength have been extracted to be plotted in Figure \ref{fig:intensitycut}. }\label{fig:nicefig}
\end{figure*}

Recently, an interesting technique to measure coronal fields with CME-driven shocks was proposed by \citet{gopalyashi2011}. This technique takes advantage of the relationship derived by \citet{russel2002} between the standoff distance of an interplanetary shock and the radius of curvature of its driver, and is applied to derive the strength of coronal fields just above the shock nose during its propagation. This technique has been applied to images obtained from white light coronagraphic observations and, recently, to CME-driven shocks observed with EUV disk imagers \citep{gopalswamy2012b} and white light heliospheric images \citep{poomvises2012} allowing for the first time the derivation of magnetic field strengths up to an heliocentric distance of $\sim 200$~\rsun. Notwithstanding the above, this technique has some limitations, in particular: 1) it can be only applied to shocks driven by CMEs, and 2) it is able to provide magnetic field measurements only along the radial located at the position of the shock nose.

\begin{figure}
\centering
\includegraphics[width=\columnwidth]{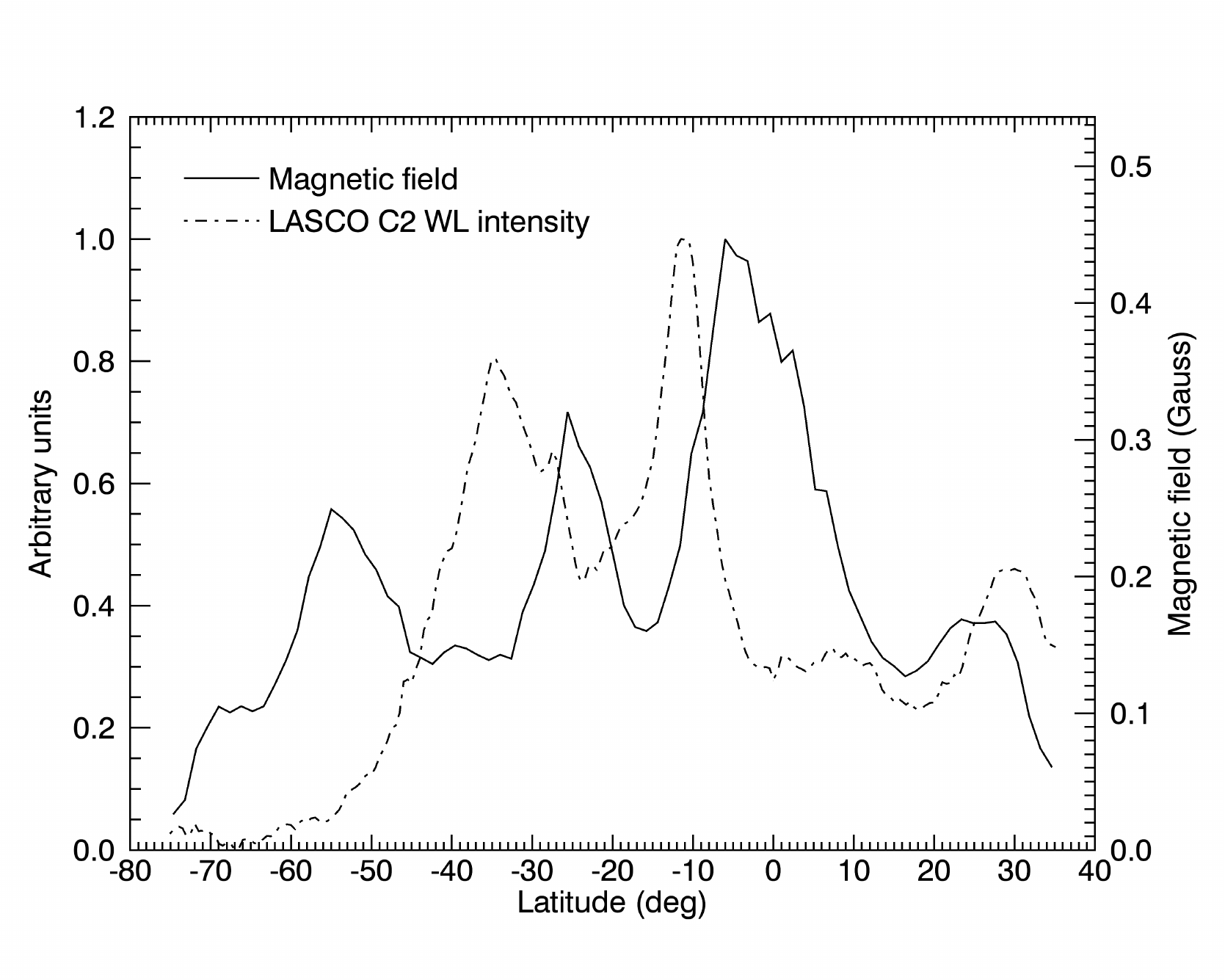}
\caption{Comparison between the normalized pre-shock coronal white light structures observed by LASCO C2 coronagraph (dashed line) and the magnetic field strengths (solid line) at the constant altitude of 2.75 R$_\odot$.}\label{fig:intensitycut}
\end{figure}

On the other hand, the technique we developed here and in our previous works is able to provide measurements of the pre-shock coronal magnetic field strengths from white light observations of shock waves over all altitudes and latitudes crossed by the shock, independently of any hypothesis on the nature of the shock driver. In fact, once a 2D map for the Alfv\'en speed and for the electron density $n_e$ are derived, the determination of the 2D coronal magnetic field strength is straightforward and is given by $B = v_\text{A} \sqrt{4 \pi n_e m_p}$. The resulting 2D map of the magnetic field strength is shown in Figure \ref{fig:fastslowwind} (top left panel) under the assumption that the solar wind speed is negligible with respect to the shock speed. Nevertheless, because the shock speed is decreasing with altitude ($v_\text{sh}\simeq 1200$~km~s$^{-1}$ at 2.5~\rsun\ and $v_\text{sh}\simeq 1000$~km~s$^{-1}$ at 12~\rsun\ as we measured at the shock center), while the wind speed is increasing, higher up in the corona the field will be more and more overestimated, leading to larger uncertainties. In order to quantify these uncertainties, lower limit estimates for the Alfv\'en speed, and thus for the magnetic field, have been derived by assuming that the whole corona is pervaded at all latitudes by fast solar wind; in particular, here we assumed the fast solar wind radial profile provided by \citet{hu1997}. The resulting 2D map for the lower limit estimate of the magnetic field strength is shown in Figure \ref{fig:fastslowwind} (bottom left panel). Comparison between the two maps clearly shows that no significant differences are present in the lower corona, while larger differences may exist higher up. In particular, by averaging all the magnetic field radial profiles obtained at different latitudes, we conclude that the maximum difference between the upper and the lower limit estimates is on the order of a factor $\sim 2.7$ at 12~\rsun, and smaller factors at lower altitudes (see Figure \ref{fig:fastslowwind}, right panel).

The magnetic field values we derived here are in very good agreement with previous measurements provided in the literature at different altitudes and latitudes and obtained with many different techniques, as shown in Figure \ref{fig:bmag_literature}. Hence, not only the radial variation of the field strength is comparable to other estimates obtained with completely different techniques, but the latitudinal modulation we derived in this work is reliable as well. We remind that the technique applied in this work for the determination of field strengths was only based on the analysis of white light coronagraphic images, which have been analyzed to derive 2D maps (projected on the plane of the sky) of the pre-shock coronal densities, shock compression ratios, shock velocities and inclination of the shock surface with respect to the radial. Then, some assumptions were needed in order to derive the magnetic field strengths: first, we assumed that above the lower boundary of the LASCO C2 occulter ($\sim 2$~\rsun) the coronal field is radial, so that the shock inclination with respect to the radial also provides its inclination with respect to the upstream magnetic field. This is not a strong assumption, because it is well known that coronal structures (outlining the magnetic field orientation) are nearly radial above heliocentric distances of $\sim 2$~\rsun. Second, we assumed an empirical formula for the determination of the Alfv\'enic Mach number for the general case of an oblique shock starting from the measured shock compression ratios and shock inclination angles. The validity of this formula has been verified in a previous work \citep{bem14} where the Alfv\'enic Mach number was derived independently also form the analysis of white light and UV data; the verification of the same formula with MHD numerical simulations has been also recently provided by another work \citep{bacchini2015}. Third, in order to convert the derived Alfv\'enic Mach numbers in estimates for the Alfv\'en speed, we assumed that the solar wind speed ahead of the shock is negligible with respect to the shock speed; as discussed above, this leads to an overestimate of the magnetic field by a factor no more than $\sim 2.7$ at 12~\rsun, decreasing with altitude. For comparison with the white light coronal structures, the magnetic field values derived in this work are shown again in Figure \ref{fig:nicefig}, plotted in the field of view of the LASCO C2 coronagraph (right panel), together with the original pre-CME coronal white light intensity (left panel). We also notice that the latitudinal distribution of coronal field strength is, in first approximation, anti-correlated with the white light intensity. This result is also better shown in Figure \ref{fig:intensitycut} providing the latitudinal distribution of the normalized WL intensity and the magnetic field strength at a constant altitude of 2.75 R$_\odot$. The observed anti-correlation is in nice agreement with what we could expect around the vertical axis of each coronal streamer, where the neutral current sheet corresponds to a region of minimum magnetic field strength.

\begin{figure*}
\centering
\vspace{-3cm}
\includegraphics[width=0.85\textwidth]{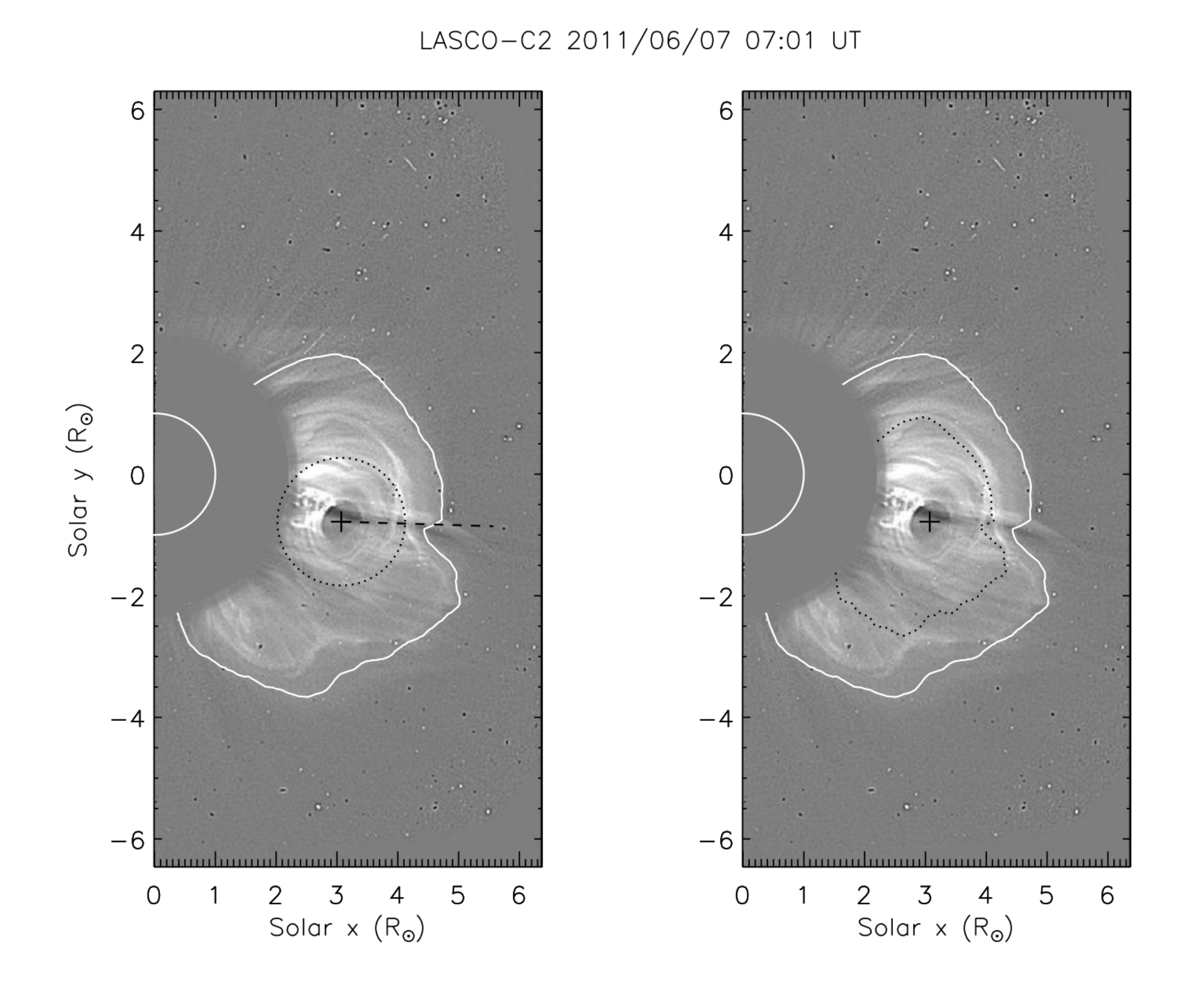}
\vspace{-0.5cm}
\caption{Left: LASCO C2 base difference image acquired on June 7, 2011 at 07:01 UT and with the contrast of faint features enhanced using the filter provided by the JHelioviewer software. The overplot shows the location of the shock (solid white line), the center of the CME flux rope (plus symbol), and the CME propagation direction (dashed black line). Right: same frame shown in the left panel, where the overplot provides again the location of the shock (solid white line), the center of the CME flux rope (plus symbol), and the location of the shock driver (black dotted line) as derived by assuming that the relationship between the Mach number at the shock nose and the $\Delta/R$ ratio holds also at different latitudes away from the shock nose (see text).}\label{fig:shock_driver}
\end{figure*}

In order to further support the correctness of our measurements of coronal magnetic fields, we also applied the same technique proposed by \citet{gopalyashi2011} and based on the measurement of the shock standoff distance. In order to perform the comparison between the two techniques, we selected the LASCO C2 frame where the circular shape of the CME flux rope is better visible, shown in Figure~\ref{fig:shock_driver}. For this frame we determined the position of the center of the flux rope (plus symbol in the left plot) and (looking at previous and subsequent frames) the CME propagation direction (dashed line in the left plot). This provides us with the identification of the shock nose, as well as a measurement of the sum between the shock standoff distance $\Delta$ and the radius $R$ of the flux rope, which turns out to be $\Delta + R = 1.48$ R$_\odot$. We thus used the value of the Mach number derived as decribed above at the shock nose ($M_A = 1.50$) and derived the expected $\Delta / R$ ratio, which turns out to be \citep[see][]{gopalyashi2011}
\begin{equation}
\frac{\Delta}{R} = K \frac{\left( \gamma -1 \right)\, M_A^2 +2}{\left( \gamma +1 \right)\, M_A^2} \simeq 0.45,
\end{equation}
where $K = 0.78$ for a circular shape of the shock driver, and $\gamma = 5/3$. With the above numbers it turns out that $\Delta = 0.46$ R$_\odot$ and $R = 1.02$ R$_\odot$. The corresponding circumference (plotted in the left panel of Figure~\ref{fig:shock_driver}) shows a quite nice agreement with the location of the CME flux rope, thus demonstrating that our results are in good agreement with those that could be derived for the same event with the technique described by \citet{gopalyashi2011}. Moreover, since in this work we derived measurements of the shock Mach number $M_A$ not only at the shock nose, but also at different latitudes, it is interesting to test what happens by assuming that the above relationship relating $M_A$ and the $\Delta / R$ ratio holds also away from the shock nose. In particular, the right plot of Figure~\ref{fig:shock_driver} shows the locations of the shock driver (black dotted line) as inferred by assuming different values of $M_A$ away from the shock nose along each radial starting from the same position of the center of the flux rope (plus symbol). The resulting curve shows a surprisingly nice agreement with some white light features visible between the CME flux rope and shock. This may suggest that at this time a decoupling between the flux rope and the shock is already occurring away from the shock nose, or alternatively that the side parts of shock are driven at some latitudes by the expansion of other loop-like plasma features surrounding the CME flux rope and embedded within the same CME.

The analysis performed here provides not only a new technique to derive coronal field strengths with unprecedent radial and latitudinal extension, but also very important insights into the physical relation between the type-II emitting regions and the shock front. In fact, the difference between the 2D maps we derived for the shock and the Alv\'en speed clearly show that in the early phases (2--4~\rsun) the whole shock surface is super-Alfv\'enic, while later on (i.e. higher up) becomes super-Alfvenic only at the nose. For a better understanding of the acceleration regions of SEP, this result has also to be considered  together with our previous finding that in the early propagation phases shocks are super-critical only at the nose and becomes sub-critical later on \citep[e.g.][]{bem11}. At the same time, we demonstrate here with analysis of radio dynamic spectra that the emission near the front was generated later than the one produced by the flanks, in agreement with the conclusion we derived from the analysis of white light data. This suggests that the acceleration of SEP leading to gradual events could also involve at different times coronal regions located not only at different altitudes, but also at different latitudes and/or longitudes along the shock front, as recently simulated for instance by \citet{rodriguez2014}.

\end{document}